\documentclass[12pt]{article}

\usepackage{amssymb,amsmath,epsfig,cite}

\normalbaselines

\begin{document}

\title{Higher-order supersymmetric quantum mechanics}

\author{David J. Fern\'andez C. and Nicol\'as Fern\'andez-Garc\'{\i}a
\\ Departamento de F\'{\i}sica, CINVESTAV \\ 
AP 14-740, 07000 M\'exico DF, Mexico}

\date{}

\maketitle

\begin{abstract}
We review the higher-order supersymmetric quantum mechanics (H-SUSY QM), 
which involves differential intertwining operators of order greater than 
one. The iterations of first-order SUSY transformations are used to derive 
in a simple way the higher-order case. The second order technique is 
addressed directly, and through this approach unexpected possibilities for 
designing spectra are uncovered. The formalism is applied to the harmonic 
oscillator: the corresponding H-SUSY partner Hamiltonians are ruled by 
polynomial Heisenberg algebras which allow a straight construction of the 
coherent states.
\end{abstract}

\newpage

\tableofcontents

\newpage

\section{Introduction}

The number of exactly solvable problems in non-relativistic quantum 
mechanics is small, and most of them can be dealt with the factorization 
method. This technique, introduced long ago by Schr\"odinger 
\cite{sc40a,sc40b}, was analyzed in depth by Infeld and Hull \cite{ih51}, 
who made an exhaustive classification of factorizable potentials. Later 
on, Witten noticed the possibility of arranging the Schr\"odinger's 
Hamiltonians into isospectral pairs (supersymmetric partners) \cite{wi81}. 
The resulting {\it supersymmetric quantum mechanics} (SUSY QM) revived the 
study of exactly solvable Hamiltonians (see e.g. \cite{uh83}). An 
additional step was Mielnik's factorization through which the general SUSY 
partner for the oscillator for a certain factorization energy was found 
\cite{mi84}; this technique was immediately applied to the hydrogen 
potential \cite{fe84}. Meanwhile Nieto \cite{ni84}, Andrianov {\it et al} 
\cite{abi84} and Sukumar \cite{su85a,su85b,su85c,su86,su87} put the method 
on its natural background discovering the links between SUSY, 
factorization and Darboux algorithm, causing then a renaissance of the 
related algebraic methods \cite{fw85,do87,bdh87,bdh88,ad88,dr88, 
bo88,dr89,le89,st89,ar90,mmz90,cflu91,la91,lr91,abcd91,cn91, 
dr93,dr95,mqs95,cln95,cks95,rr95,fno96,pe96,ca96,rso96,cz97, 
jr97,jr98,cmpr98,mnn98,mpog99,cr99,cr00,nnr00,ba01,li02, 
nps03,mgqgg03,srg03} (for a recent review see \cite{mr04} and references 
therein).

These procedures can be recovered from a general setting in which a 
first-order differential operator intertwines two Hamiltonians 
\cite{dc78,ca79}. This so-called first-order intertwining technique 
suggests further generalizations: the most obvious one involves a $k$-th 
order differential intertwining operator and gives place to the 
higher-order supersymmetric quantum mechanics (H-SUSY QM) 
\cite{ais93,aicd95,bs97,fe97,fgn98,fhm98,ro98a,ro98b,fh99, 
bgbm99,mnr00,pl00,fr01,anst01,sk03,pl04,cfnn04,ac04,fr04}. It can be 
achieved by iterations of first order SUSY transformations \cite{fhm98}. 
An alternative to deal with the problem consists in looking for the $k$-th 
order operator directly, expressing the intertwiner as a sum of the $k+1$ 
terms $g_i(x)d^i/dx^i, i=0,\dots, k$, and solving the system of equations 
resulting from the intertwining relationship for the $g_i(x)$'s. By 
assuming that the initial Hamiltonian is solvable, new solvable ones and 
their eigenstates can be generated through the two previous formulations.

Concerning applications, the procedure works successfully to generate 
$k$-parametric families of potentials almost isospectral to the harmonic 
oscillator \cite{fhm98,fh99}, the radial hydrogen-like potentials 
\cite{ro98a,ro98b}, and in the free particle case \cite{mnr00} (for a 
collection of recent papers on SUSY QM see \cite{afhnns04}). Specially 
interesting are the SUSY partners of the oscillator because through them 
some connections with other important subjects of mathematical physics can 
be established.

In the first place let us notice the existence of differential 
annihilation and creation operators of order greater than one for the SUSY 
partners of the oscillator. These operators provide natural realizations 
of the so called polynomial Heisenberg algebras 
\cite{fh99,cfnn04,dek92,ad94,ek95,as97,aaw99}. It is interesting as well 
that these non-linear algebras admit partial linearizations 
\cite{fhn94,fnr95,ro96}, i.e., through appropriate modifications on the 
ladder operators the standard Heisenberg algebra is recovered on the 
subspace spanned by the eigenfunctions intertwined with the physical 
eigenstates of the oscillator.

In the second place let us mention the construction of coherent states 
(CS) for potentials generated through the intertwining technique 
\cite{fh99,as97,fhn94,fnr95,ro96,fa93,sp95,kk96,fs96,bs96,ek97,sbl98, 
cjt98}. In this direction Fukui and Aizawa have derived some CS for the 
`shape invariant' potentials \cite{fa93}, a particular class of solvable 
potentials generated long ago by Infeld and Hull through the factorization 
method \cite{ih51}. The first work involving CS for the simplest 
non-trivial family of potentials isospectral to the oscillator 
(Abraham-Moses \cite{mi84,am80}) was done in 1994 \cite{fhn94} (see also 
\cite{fnr95,ro96}). Later on various developments have appeared, e.g., 
some authors have constructed the CS for a class of anharmonic oscillators 
which spectra consist of a part isospectral to the oscillator plus one 
lower energy at a multiple of the spacing between the oscillator levels 
\cite{bs96}. Furthermore, CS have been derived \cite{as97} for families of 
SUSY partner potentials having an arbitrary eigenvalue below the ground 
state energy of the oscillator \cite{su85b,jr97,fhm98}. Recently, it has 
been implemented as well the CS construction for the general H-SUSY 
partners of the oscillator \cite{fh99}.

In these lecture notes we are going to address the subjects mentioned 
above. We will start with the iterative H-SUSY QM to generate families of 
potentials (almost) isospectral to the initial one. Then, we will 
formulate the same problem directly, by assuming that the intertwining 
operator is of $k$-th order. Due to the difficulty involved in this 
problem, we just will illustrate the technique through the second-order 
supersymmetric quantum mechanics. Then we will apply these generation 
procedures to the harmonic oscillator. It will be shown that the intrinsic 
symmetry of the H-SUSY partners of the oscillator is generated by a pair 
of differential annihilation and creation operators of order greater than 
one, giving place to explicit representations of the polynomial Heisenberg 
algebras. We will show as well an interesting linearization procedure for 
the nonlinear algebras characteristic of the H-SUSY partners of the 
oscillator. We will perform then the coherent states construction in the 
non-linear as well as in the linear cases. We will conclude these notes 
with some discussion on the future of SUSY QM.

\newpage

\section{Higher-order supersymmetric quantum mechanics: iterative 
approach}

Let us start with two Schr\"odinger Hamiltonians
\begin{equation}
H_i = - \frac12\frac{d^2}{dx^2} + V_i(x) \qquad
i=0,1
\end{equation}
and suppose the existence of a first order differential operator 
$A_1^\dagger$ intertwining them
\begin{equation}
H_1 A_1^\dagger = A_1^\dagger H_0 \label{intertwining1}
\end{equation}
where
\begin{equation}
A_1^\dagger= \frac{1}{\sqrt{2}}\left( -\frac{d}{dx} + 
\alpha_1(x)\right)
\end{equation}
the superpotential $\alpha_1(x)$ being a real function to be determined. 
In order to `pass' the differential operator $d/dx$ and its powers to the 
right in the terms arising from (\ref{intertwining1}), we use the 
following operator relationships
\begin{eqnarray} 
&& \frac{d}{dx}f = f\frac{d}{dx} + f' \nonumber \\ 
[10pt] && \frac{d^2}{dx^2}f = f\frac{d^2}{dx^2} + 2 f'\frac{d}{dx} + f'' 
\nonumber \end{eqnarray} 
 Then, it is straightforward to show that
\begin{eqnarray} 
&& \sqrt{2}H_1 A_1^\dagger = \frac12\frac{d^3}{dx^3} -\frac{\alpha_1}{2} 
\frac{d^2}{dx^2} - (V_1+\alpha_1') \frac{d}{dx} + \alpha_1 V_1 - 
\frac{\alpha_1''}{2} \nonumber \\
&& \sqrt{2}A_1^\dagger H_0 = \frac12\frac{d^3}{dx^3} -
\frac{\alpha_1}{2} \frac{d^2}{dx^2} - V_0 \frac{d}{dx} + \alpha_1 V_0 - V_0' 
\nonumber 
\end{eqnarray} 
Due to (\ref{intertwining1}) we make equal the coefficient of the same 
powers of $d/dx$ of the two previous equations to obtain
\begin{eqnarray} 
&& V_1 = V_0 - \alpha_1' \label{nV} \\ 
&& \alpha_1 V_1 - \frac{\alpha_1''}{2} = \alpha_1 V_0 - V_0' 
\label{riccatiprime} 
\end{eqnarray} 
Substituting the expression for $V_1$ of (\ref{nV}) in (\ref{riccatiprime}) 
and integrating the result we get:
\begin{equation}
\alpha_1' + \alpha_1^2 = 2(V_0 - \epsilon) \label{riccati1}
\end{equation}
From now on we are going to express explicitly the dependence of the 
superpotential in terms of the {\it factorization energy} $\epsilon$ in 
the way $\alpha_1(x,\epsilon)$. If a function $u^{(0)}(x)$ such that 
$\alpha_1(x,\epsilon)= {u^{(0)}}'/u^{(0)}$ is used, we have that 
(\ref{nV},\ref{riccati1}) become:
\begin{eqnarray}
&& -\frac12 {u^{(0)}}'' + V_0 u^{(0)} = \epsilon u^{(0)}
\label{schrodinger0}\\
&& V_1 = V_0 - \left( \frac{{u^{(0)}}'}{u^{(0)}}  \right)'  \label{nVu}
\end{eqnarray}
i.e., $u^{(0)}$ is a solution (not necessarily physical) of the initial 
stationary Schr\"odinger equation associated to $\epsilon$.

Let us notice that (\ref{nV},\ref{riccati1}) guarantee that $H_0$ and 
$H_1$ become factorized:
\begin{equation}
H_0 = A_1 A_1^\dagger + \epsilon \qquad H_1 = A_1^\dagger A_1 +
\epsilon
\end{equation}
where
\begin{equation}
A_1 = \frac{1}{\sqrt{2}}\left(\frac{d}{dx} + \alpha_1(x,\epsilon)\right)
\end{equation}
is the operator adjoint to $A_1^\dagger$.

Suppose that $V_0(x)$ is a solvable potential with eigenfunctions 
$\psi_n^{(0)}(x)$ and eigenvalues $E_n, \ n=0,1,\dots$ Furthermore, let us 
assume that we have found a solution $\alpha_1(x,\epsilon_1)$ 
($u^{(0)}_1(x)$) to the Riccati equation (\ref{riccati1}) (Schr\"odinger 
equation (\ref{schrodinger0})) for a given value of the factorization 
energy $\epsilon = \epsilon_1 \leq E_0$, where $E_0$ is the ground state 
energy of $H_0$. Thus, the $V_1(x)$ given in (\ref{nV}) is a completely 
determined solvable potential with normalized eigenfunctions
\begin{eqnarray}
&& \psi_{\epsilon_1}^{(1)}(x) \propto e^{-\int_0^x\alpha_1(y,
\epsilon_1) dy}= \frac{1}{u^{(0)}_1(x)} \label{missing1} \\ 
&& \psi_n^{(1)}(x) =
\frac{A_1^\dagger\psi_n^{(0)}(x)}{\sqrt{E_n-\epsilon_1}}   \label{npsi}
\end{eqnarray}
and eigenvalues $\{\epsilon_1, E_n, \ n=0,1,\dots\}$. Let us remark that 
the restriction $\epsilon_1\leq E_0$ is crucial to avoid the existence of 
singularities in $\alpha_1(x,\epsilon_1)$, in $V_1$ and also in the 
$\psi_{\epsilon_1}^{(1)}, \ \psi_n^{(1)}$ of (\ref{missing1},\ref{npsi}). 
Indeed, if $\epsilon_1$ would be greater than $E_0$, the transformation 
function $u_1^{(0)}$ would have always zeros in the initial $x$-domain and 
thus $\alpha_1(x,\epsilon_1)$ would have singularities at those points. If 
however $\epsilon_1 \leq E_0$, then $u^{(0)}_1(x)$ will have at most one 
zero; by exploring the two-dimensional subspace of solutions associated to 
$\epsilon_1$ it is possible to find a subset of nodeless solutions (see 
e.g. \cite{su85a,su85b}). By simplicity, we shall assume that the 
factorization energy used to generate any new Hamiltonian through the 
first-order SUSY is below the ground state energy of the initial 
Hamiltonian. We shall suppose as well that, for fixed $\epsilon_1$, the 
arbitrary parameter of a general Riccati solution (\ref{riccati1}) has 
been adjusted in order to avoid the singularities in the $\alpha$'s.

Now we iterate the previous technique, taking $V_1(x)$ as the given 
solvable potential to generate a new one $V_2(x)$ through an intertwining 
operator $A_2^\dagger$ and a different factorization energy $\epsilon_2$, 
with $\epsilon_2<\epsilon_1$. The corresponding intertwining relationship, 
$H_2 A_2^\dagger = A_2^\dagger H_1$, leads to equations similar to 
(\ref{nV},\ref{riccati1}):
\begin{eqnarray}
&& \alpha_2'(x,\epsilon_2) + \alpha_2^2(x,\epsilon_2) = 2[V_1(x) - 
\epsilon_2] \label{riccati2} \\
&& V_2(x) = V_1(x) - \alpha_2'(x,\epsilon_2) \label{nV2}
\end{eqnarray}
In terms of transformation functions $u_2^{(1)}(x)$ such that 
$\alpha_2(x,\epsilon_2)= {u_2^{(1)}}'/u_2^{(1)}$ we have:
\begin{eqnarray}
&& -\frac12 {u_2^{(1)}}'' + V_1 u_2^{(1)} = \epsilon_2 u_2^{(1)}
\label{schrodinger1}\\
&& V_2 = V_1 - \left( \frac{{u_2^{(1)}}'}{u_2^{(1)}}  \right)' \label{nVu1}
\end{eqnarray}

An important result to be proved now is that the solutions to 
(\ref{riccati2}) can be algebraically determined through solutions of the 
initial Riccati equation (\ref{riccati1}) for the factorization energies 
$\epsilon_1$ and $\epsilon_2$ \cite{fhm98,ro98a,ro98b,fh99,bgbm99, mnr00}. 
We stick to \cite{fg04} because there the calculations are transparent. To 
find the corresponding formula \cite{mnr00}, we know that the two initial 
Riccati solutions satisfy:
\begin{eqnarray}
\alpha'_1(x,\epsilon_1) + \alpha^2_1(x,\epsilon_1)=2[V_{0}(x) - 
\epsilon_1] \nonumber \\
\alpha'_1(x,\epsilon_2) + \alpha^2_1(x,\epsilon_2)=2[V_{0}(x) - 
\epsilon_2] \label{ricattidoble}
\end{eqnarray}\\
At the Schr\"odinger level we have that 
\begin{eqnarray}
H_0 u^{(0)}_1(x)=\epsilon_1u^{(0)}_1(x) \nonumber \\
H_0 u^{(0)}_2(x)=\epsilon_2u^{(0)}_2(x)
\end{eqnarray}
where
\begin{eqnarray*}
u^{(0)}_1(x) \propto e^{\int_{0}^{x}\alpha_1(y,\epsilon_1)dy} \\
u^{(0)}_2(x) \propto e^{\int_{0}^{x}\alpha_1(y,\epsilon_2)dy} \\
\end{eqnarray*}
Let us remember that $u^{(0)}_1(x)$ is used to implement the first SUSY 
transformation and that the eigenfunction of $H_1$ associated to 
$\epsilon_1$ is given by (\ref{missing1}). On the other hand, the 
eigenfunction of $H_1$ associated to $\epsilon_2$ is given by:
\begin{equation}
u^{(1)}_2 \propto  A^\dagger_1 u^{(0)}_2 \propto 
-{u^{(0)}_2}' + \alpha_1(x,\epsilon_1)u^{(0)}_2 \propto
\frac{W(u^{(0)}_1,u^{(0)}_2)}{u^{(0)}_1}
\end{equation}
and since that
\begin{eqnarray*}
{u^{(0)}_2}' =\alpha_1(x,\epsilon_2)u^{(0)}_2
\end{eqnarray*}
we get:
\begin{equation}
u^{(1)}_2 \propto \left[\alpha_1(x,\epsilon_1)-
\alpha_1(x,\epsilon_2)\right] u^{(0)}_2
\label{2.11}
\end{equation}
In order to implement the second SUSY transformation, we express 
$u^{(1)}_2$ in the standard way in terms of the corresponding 
superpotential:
\begin{equation}
u^{(1)}_2(x)\propto e^{\int_{0}^{x}\alpha_2(y,\epsilon_2)dy}
\label{2.12}
\end{equation}
By plugging (\ref{2.12}) in (\ref{2.11}) we arrive at
\begin{equation}
e^{\int_{0}^{x}\alpha_2(y,\epsilon_2)dy} \propto
\left[\alpha_1(x,\epsilon_1)-\alpha_1(x,\epsilon_2)\right] u^{(0)}_2
\end{equation}
By taking the logarithm of both sides of the previous equation:
\begin{equation}
\int_{0}^{x}\alpha_2(y,\epsilon_2)dy = \ln u^{(0)}_2 + \ln
\left[\alpha_1(x,\epsilon_1)-\alpha_1(x,\epsilon_2)\right]+{\rm constant}
\end{equation}
Deriving this expression with respect to $x$:
\begin{eqnarray}
&& \alpha_2(x,\epsilon_2) =  
\alpha_1(x,\epsilon_2)+\frac{\alpha_1'(x,\epsilon_1)-
\alpha_1'(x,\epsilon_2)}{\alpha_1(x,\epsilon_1)-\alpha_1(x,\epsilon_2)}
\end{eqnarray}
By using the initial Riccati equations (\ref{ricattidoble}) we obtain:
\begin{equation}
\frac{\alpha'_1(x,\epsilon_1) - \alpha'_1(x,\epsilon_2)}{
\alpha_1(x,\epsilon_1) - \alpha_1(x,\epsilon_2)} = 
- \alpha_1(x,\epsilon_2)-\alpha_1(x,\epsilon_1) -
\frac{2(\epsilon_1 - \epsilon_2)}{\alpha_1(x,\epsilon_1) - 
\alpha_1(x,\epsilon_2)}
\end{equation}
Therefore
\begin{equation}
\alpha_2(x,\epsilon_2) = - \alpha_1(x,\epsilon_1) -
\frac{2(\epsilon_1 - \epsilon_2)}{\alpha_1(x,\epsilon_1) - 
\alpha_1(x,\epsilon_2)}
\label{backlund1}
\end{equation}
This formula expresses the solution to (\ref{riccati2}) with $V_1(x) = 
V_0(x)-\alpha_1'(x,\epsilon_1)$ in form of a finite difference formula 
involving two solutions $\alpha_1(x,\epsilon_1)$, $\alpha_1(x,\epsilon_2)$ 
of the Riccati equation (\ref{riccati1}) for the factorization energies 
$\epsilon_1, \ \epsilon_2$ (see also \cite{fhm98}). Notice that a similar 
formula has been used by Adler in order to discuss the B\"acklund 
transformations of the Painlev\'e equations \cite{ad94}. The potential 
$V_2(x)$ reads:
\begin{equation}
V_2(x) = V_1(x) - \alpha'_2(x,\epsilon_2) = V_0(x) + 
\left[\frac{2(\epsilon_1 - \epsilon_2)}{\alpha_1(x,\epsilon_1) - 
\alpha_1(x,\epsilon_2)}\right]'
\end{equation}
The eigenfunctions associated to $H_2$ are given by:
\begin{eqnarray}
&& \psi_{\epsilon_2}^{(2)}(x) \propto e^{-\int_0^x \alpha_2(y, 
\epsilon_2) dy} = \frac{1}{u_2^{(1)}} \nonumber \\
&& \psi_{\epsilon_1}^{(2)}(x) =     
\frac{A_2^\dagger
\psi_{\epsilon_1}^{(1)}(x)}{\sqrt{\epsilon_1-\epsilon_2}} \label{eigenh2} 
\\ && \psi_n^{(2)}(x) =\frac{A_2^\dagger\psi_n^{(1)}(x)}{\sqrt{E_n - 
\epsilon_2}} = \frac{A_2^\dagger 
A_1^\dagger\psi_n^{(0)}(x)}{\sqrt{(E_n - \epsilon_1
)(E_n - \epsilon_2)}} \nonumber
\end{eqnarray}  
The corresponding eigenvalues are $\{\epsilon_2, \ \epsilon_1, \ E_n, \
n=0,1,\dots\}$.

We can continue the iteration process as many times as solutions for 
different values $\epsilon_i$ to the initial Riccati equation 
(\ref{riccati1}) we have. Suppose that we know $k$ of these, $\{ 
\alpha_1(x,\epsilon_i), \ i=1,\dots,k, \ \epsilon_{i+1}<\epsilon_i\}$, and 
we iterate the process $k$ times. Hence, a new solvable Hamiltonian $H_k$ 
will be gotten whose potential reads:
\begin{equation}
V_k(x) = V_{k-1}(x) - \alpha_k'(x,\epsilon_k) = V_0(x) - \sum_{i=1}^k  
\alpha_i'(x,\epsilon_i) \label{nVk}
\end{equation}
where $\alpha_i(x,\epsilon_i)$ is given by a recursive finite difference 
formula generalizing (\ref{backlund1}):
\begin{equation}
\alpha_{i+1}(x,\epsilon_{i+1}) = - \alpha_{i}(x,\epsilon_{i}) - 
\frac{2(\epsilon_{i}-\epsilon_{i+1})}{\alpha_{i}(x,\epsilon_{i}) -
\alpha_{i}(x,\epsilon_{i+1})} \qquad i=1,\dots,k-1 \label{backlundi}
\end{equation}  
The eigenfunctions are given by:
\begin{eqnarray}
\psi_{\epsilon_k}^{(k)}(x) \propto & e^{- \int_0^x \alpha_k(y,
\epsilon_k) dy} \nonumber \\
\psi_{\epsilon_{k-1}}^{(k)}(x) = &
\frac{A_k^\dagger \psi_{\epsilon_{k-1}}^{(k-1)} (x)}{\sqrt{\epsilon_{k-1}
- \epsilon_k}} \nonumber \\
& \vdots \label{eigenhk} \\
\psi_{\epsilon_1}^{(k)}(x) = &
\frac{A_k^\dagger\dots A_2^\dagger \psi_{\epsilon_1}^{(1)}(x)}{
\sqrt{(\epsilon_1-\epsilon_2)\dots(\epsilon_1-\epsilon_k)}} \nonumber \\
\psi_n^{(k)}(x) = & \frac{A_k^\dagger \dots A_1^\dagger\psi_n^{(0)}(x)}{
\sqrt{(E_n-\epsilon_1)\dots(E_n - \epsilon_k)}} \nonumber
\end{eqnarray}
The corresponding eigenvalues are $\{\epsilon_i, \ E_n, \ i=k,\dots,1, \ 
n=0,1,\dots\}$.

In order to have the scheme complete, let us remember how are intertwined 
the $H_i$'s:
\begin{equation}
H_i A_i^\dagger = A_i^\dagger H_{i-1} \qquad i=1,\dots,k
\end{equation}
Thus, departing from $H_0$ we have generated a chain of factorized
Hamiltonians: 
\begin{eqnarray}
& H_i = A_i^\dagger A_i + \epsilon_i = A_{i+1}A_{i+1}^\dagger
+\epsilon_{i+1} \qquad i=1,\dots ,k-1 \\
& H_k = A_k^\dagger A_k +\epsilon_k
\end{eqnarray}
where the end potential $V_k(x)$ can be recursively determined by means of 
(\ref{nVk},\ref{backlundi}). Thus, we require the mentioned $k$ solutions 
$\alpha_1(x,\epsilon_i), \ i=1,\dots,k$ to the initial Riccati equation 
(\ref{riccati1}), which means to have $k$ non-equivalent factorizations of 
the Hamiltonian $H_0$:
\begin{equation}
H_0 = \frac12 \left(\frac{d}{dx} + \alpha_1(x,\epsilon_i) \right)
\left(-\frac{d}{dx} + \alpha_1(x,\epsilon_i) \right) + \epsilon_i \qquad
i=1,\dots,k
\end{equation}
Let us notice that there is a $k$-th order differential operator, 
$B_k^\dagger = A_k^\dagger \dots A_1^\dagger$, intertwining the initial 
$H_0$ and final Hamiltonians $H_k$:
\begin{equation}  
H_k B_k^\dagger = B_k^\dagger H_0 \label{intertwiningk}
\end{equation}
From equations (\ref{eigenhk}) we get:
\begin{equation}
B_k^\dagger \psi_n^{(0)} = \sqrt{(E_n-\epsilon_1)\dots(E_n-\epsilon_k)}
\,\psi_n^{(k)}
\end{equation}
while from the adjoint to (\ref{intertwiningk}) it turns out that:
\begin{equation}
B_k\psi_n^{(k)} = \sqrt{(E_n-\epsilon_1)\dots(E_n-\epsilon_k)}
\,\psi_n^{(0)}
\end{equation}
These equations lead to the higher-order supersymmetric quantum mechanics 
H-SUSY QM 
\cite{ais93,aicd95,bs97,fe97,fgn98,fhm98,ro98a,ro98b,fh99,bgbm99, mnr00}. 
In this treatment the standard SUSY algebra with two generators 
\cite{wi81}
\begin{equation}
[Q_i, H_{\rm ss}]=0 \qquad \{ Q_i,Q_j\} = \delta_{ij} H_{\rm ss} \qquad
i,j=1,2 \label{susyalg}
\end{equation}
is realized with the aid of $B_k$ and $B_k^\dagger$:
\begin{eqnarray}
& Q = \left(\begin{matrix} 0 & 0 \cr B_k & 0 \end{matrix} \right) \qquad
Q^\dagger = \left(\begin{matrix} 0 & B_k^\dagger \cr 0 & 0 \end{matrix}
\right) \\
& H_{\rm ss} = \{ Q, Q^\dagger\} = \left(\begin{matrix} B_k^\dagger B_k & 
0
\cr 0 & B_k B_k^\dagger \end{matrix} \right) 
\end{eqnarray}
where $Q_1 = (Q^\dagger + Q)/\sqrt{2}, \ Q_2 = i (Q - 
Q^\dagger)/\sqrt{2}$. Due to
\begin{eqnarray}
B_k^\dagger B_k & = & A_k^\dagger \dots A_1^\dagger A_1 \dots A_k = 
A_k^\dagger \dots A_2^\dagger (H_1-\epsilon_1)A_2 \dots A_k \nonumber \\ 
& = & A_k^\dagger \dots A_3^\dagger(H_2-\epsilon_1)(H_2-\epsilon_2)
A_3 \dots A_k
= (H_k-\epsilon_1)\dots(H_k-\epsilon_k) \nonumber \\
B_k B_k^\dagger & = & A_1 \dots A_k A_k^\dagger \dots  A_1^\dagger = 
A_1 \dots A_{k-1}(H_{k-1}-\epsilon_k) A_{k-1}^\dagger \dots A_1^\dagger
\nonumber \\
&=&(H_0-\epsilon_1)\dots(H_0-\epsilon_k) \nonumber
\end{eqnarray}
it turns out that the SUSY generator $H_{\rm ss}$ is a $k$-th order 
polynomial
\begin{equation}
H_{\rm ss} = (H_{\rm s}^p-\epsilon_1) \dots (H_{\rm s}^p-\epsilon_k)
\end{equation}
of the physical Hamiltonian $H_{\rm s}^p$ involving the intertwined 
Hamiltonians $H_0$ and $H_k$
\begin{equation}
H_{\rm s}^p = \left(\begin{matrix} H_k & 0 \cr 0 & H_0 \end{matrix} 
\right) \label{hssp}
\end{equation}
In particular, if $k=1$ we will get the standard SUSY QM for which $H_{\rm 
ss} = (H_{\rm s}^p-\epsilon_1)$, i.e., $H_{\rm ss}$ is linear in $H_{\rm 
s}^p$. If $k=2$ we will get the quadratic superalgebra, or SUSUSY QM due 
to $H_{\rm ss} = (H_{\rm s}^p-\epsilon_1)(H_{\rm s}^p-\epsilon_2)$ 
\cite{ais93,fe97,fgn98}.

\newpage

\section{Higher-order supersymmetric quantum mechanics: direct approach}

In the direct procedure to the H-SUSY QM it is supposed from the very 
beginning that the intertwining operator in (\ref{intertwiningk}) is of 
$k$-th order:
\begin{equation}
B_k^\dagger = 2^{-k/2}\left[ (-1)^k \frac{d^k}{dx^k} + g_{k-1}(x) 
\frac{d^{k-1}}{dx^{k-1}} + \dots + g_1(x) \frac{d}{dx} + g_0(x)
\right]
\end{equation}
where the real functions $\{ g_i(x), i=0,\dots,k-1\}$ can in principle be 
determined through an approach similar to the one followed in the 
first-order case. Let us notice that the formulae 
(\ref{intertwiningk}-\ref{hssp}) are still valid in this case, but the 
procedure to derive them is long and involved. Here we present just the 
simplest case with $k=2$. This will illustrate clearly the advantages to 
design spectra supplied by the direct procedure compared with the 
iterative one.

\subsection{Second-order supersymmetric quantum mechanics}

The second-order supersymmetric quantum mechanics 
\cite{ais93,aicd95,bs97,fe97,fr04,fg04} arises when considering a 
second order intertwining operator such that
\begin{eqnarray}
&& H_2 B_2^\dagger = B_2^\dagger H_0  \label{intertwining2} \\
&& H_i=- \frac12 \frac{d^2}{dx^2} + V_i(x) \ \qquad i=0,2 \nonumber 
\\
&& B_2^\dagger = \frac12\left(\frac{d^2}{dx^2} - \eta(x)\frac{d}{dx} + 
\gamma(x)\right) \nonumber
\end{eqnarray}
The calculation of the left hand side of (\ref{intertwining2}) leads to:
\begin{eqnarray}
2 H_2B_2^\dagger
&=& -\frac{1}{2}\frac{d^4}{dx^4}+\frac{\eta}{2}\frac{d^3}{dx^3}+
\left(\eta'-\frac{\gamma}{2}+V_2\right)\frac{d^2}{dx^2} \nonumber \\
&& +\left(\frac{\eta''}{2}-\gamma'- \eta V_2\right)\frac{d}{dx}
+\gamma V_2 -\frac{\gamma''}{2} \label{2.21}
\end{eqnarray}
The corresponding right hand side provides:
\begin{eqnarray}
2 B_2^{\dagger}H_0&=&-\frac{1}{2}\frac{d^4}{dx^4}+\frac{\eta}{2}
\frac{d^3}{dx^3}+\left(V_0-\frac{\gamma}{2}\right)\frac{d^2}{dx^2} 
\nonumber \\
&& +\left(2V_0'-\eta V_0\right)\frac{d}{dx} + V_0''-\eta V_0'+\gamma V_0 
\label{2.22}
\end{eqnarray}
The coefficients of the same powers of $d/dx$ in (\ref{2.21},\ref{2.22}) 
must be equal, and thus we arrive at:
\begin{eqnarray} 
& V_2=V_0-\eta'  \label{2.23} \\
& \frac{\eta''}{2}-\gamma'-\eta V_2=2V_0'-\eta V_0 \label{2.24} \\
& \gamma V_2-\frac{\gamma''}{2}=V_0''-\eta V_0' +\gamma V_0
\label{2.25}\end{eqnarray}
By substituting (\ref{2.23}) in (\ref{2.24}) and solving for $\gamma'$ we 
get:
\begin{equation}
\gamma'=\frac{\eta''}{2} + \eta\eta' - 2V_0'
\label{gammaprima}
\end{equation}
Integrating this equation with respect to $x$:
\begin{equation} 
\gamma = \frac{\eta'}{2} + \frac{\eta^2}{2} - 2V_0 + d
\label{gamma}
\end{equation}
where $d$ is a real constant. By plugging (\ref{2.23}) in (\ref{2.25}) and 
solving for $\gamma''$ we have:
\begin{equation}
\gamma'' = 2\left(\eta V_0' - V_0'' - \gamma\eta'\right)
\label{gammabiprima}
\end{equation}
Taking the derivative of (\ref{gammaprima}) with respect to $x$ we get:
\begin{equation}
\gamma'' = \frac{\eta'''}{2} + \eta\eta'' + \eta'^2 - 2V_0''
\label{2.29}
\end{equation}
If we substitute (\ref{gamma}) and (\ref{2.29}) in (\ref{gammabiprima}) we 
arrive at
\begin{eqnarray*}
\frac{\eta'''}{2} + \eta\eta'' + 2\eta'^2 =
2\left(\eta V_0' + 2\eta' V_0\right) -\eta^2\eta' - 2d\eta'
\end{eqnarray*}
If we multiply the previous equation by $\eta$, then add and subtract 
$\eta'\eta''/2$ we arrive at:
\begin{eqnarray*}
2\frac{d}{dx}\left(\eta^2V_0\right)&=&\frac{\eta\eta'''}{2} +
\frac{\eta'\eta''}{2} + \frac{d}{dx}(\eta^2\eta')
+\eta^3\eta' + 2d\eta\eta' - \frac{\eta'\eta''}{2}
\end{eqnarray*}
Integrating with respect to $x$ and moving all the terms to the right hand 
side we get:
\begin{eqnarray}
&&\frac{\eta\eta''}{2}-\frac{\eta'^2}{4} +\eta^2\eta' +\frac{\eta^4}{4}
 - 2V_0 \eta^2 + d\eta^2 + c = 0 \label{2.33}
\end{eqnarray}
being $c$ a real constant. Therefore, given $V_0(x)$ the new potential 
$V_2(x)$ and the function $\gamma(x)$ are obtained through (\ref{2.23}) 
and (\ref{gamma}) once we know explicitly a solution $\eta(x)$ of 
(\ref{2.33}). To get it, we make the Ans\"atz \cite{ro98a,ro98b,fr04}
\begin{equation}
\eta' = -\eta^2 + 2\beta\eta + 2\xi \label{2.34} 
\end{equation}
where $\beta(x)$ and $\xi(x)$ are functions to be determined. Therefore:
\begin{eqnarray}
\eta'' & = & - 2\eta\eta' + 2\beta'\eta + 2\beta\eta' + 2\xi'
\label{2.35} \\
\frac{\eta\eta''}{2} & = & - \eta^2\eta' + \beta'\eta^2 +
\beta\eta\eta' + \eta\xi' \label{2.36} \\
- \frac{\eta'^2}{4} & = & - \frac{\eta^4}{4}+\beta\eta^3 +
\left(\xi - \beta^2\right)\eta^2 - 2\beta\xi\eta - \xi^2
\label{2.37}
\end{eqnarray}
The substitution of (\ref{2.36}) and (\ref{2.37}) into (\ref{2.33}) leads 
to:
\begin{eqnarray}
&\beta'\eta^2 + \beta\eta\eta' + \eta\xi'
+ \beta\eta^3 + (\xi - \beta^2)\eta^2 \nonumber \\
& - 2\beta\xi\eta - \xi^2 - 2 V_0\eta^2 + d\eta^2 + c = 0
\label{2.38}
\end{eqnarray}
and using again (\ref{2.34}) to eliminate $\eta'$ in the previous equation 
we arrive at:
\begin{equation}
\left(\beta' + \beta^2 - 2V_0 + \xi + d\right)\eta^2 + \xi'\eta - 
\xi^2 + c = 0
\label{2.39}
\end{equation}
Since (\ref{2.39}) must be valid for arbitrary $\eta$, the coefficients of 
each power of $\eta$ must vanish, which leads to $\xi^2\equiv c$. 
Therefore:
\begin{equation} 
\beta'(x)+\beta^2(x)=2[V_0(x)-\epsilon] \label{2.40}
\qquad \epsilon=\frac{d+\xi}{2}
\end{equation}
Alternatively, we can work with the Schr\"odinger equation related to 
(\ref{2.40}) through the change $\beta = {u^{(0)}}'/u^{(0)}$ \cite{crf01}:
\begin{equation}
-\frac12 {u^{(0)}}'' + V_0 u^{(0)} = \epsilon u^{(0)}
\label{schrodinger}
\end{equation}

Depending on whether $c$ is zero or not, $\xi$ vanishes or takes two 
different values $\xi= \pm\sqrt{c}$. If $c=0$, we need to solve one 
equation of form (\ref{2.40}) and then the resulting (\ref{2.34}) for 
$\eta(x)$. If $c\neq 0$ there will be two different equations of type 
(\ref{2.40}), with factorization energies $\epsilon_1 \equiv (d 
+\sqrt{c})/2$ and $\epsilon_2 \equiv (d - \sqrt{c})/2$. Once we solve 
them, it is possible to construct algebraically a common solution 
$\eta(x)$ of the corresponding pair of equations (\ref{2.34}). There is an 
obvious difference between the real case with $c>0$ and the complex case 
with $c<0$; thus, it follows a natural scheme of classification for the 
solutions $\eta(x)$ based on the sign of $c$.

\subsection{Classification of the second-order SUSY transformations}

\subsubsection{The real case with $c>0$} 

Here we have that $\epsilon_1,\epsilon_2\in{\mathbb R}$, $\epsilon_1 \neq 
\epsilon_2$, and the corresponding Riccati solutions of (\ref{2.40}) are 
denoted by $\beta_1(x)$, $\beta_2(x)$ respectively. The associated pair of 
equations (\ref{2.34}) become
\begin{eqnarray}
&& \eta'(x) = -\eta^2(x) + 2 \beta_1(x) \eta(x)+2(\epsilon_1-\epsilon_2)
\nonumber \\ 
&& \eta'(x) = -\eta^2(x) + 2 \beta_2(x) \eta(x)+2(\epsilon_2-\epsilon_1)
\nonumber   
\end{eqnarray}
By subtracting them we get algebraically $\eta(x)$ in terms of 
$\epsilon_1, \ \epsilon_2$ and $\beta_1(x), \ \beta_2(x)$:
\begin{equation}
\eta(x) =  -\frac{2(\epsilon_1 - \epsilon_2)}{\beta_1(x) -\beta_2(x)}
\label{backlund}
\end{equation}
If the corresponding Schr\"odinger solutions are used we have 
\begin{equation}
\eta(x) = 
\frac{2(\epsilon_1 - 
\epsilon_2)u^{(0)}_1 u^{(0)}_2}{W(u^{(0)}_1,u^{(0)}_2)} 
= \frac{W'(u^{(0)}_1,u^{(0)}_2)}{W(u^{(0)}_1,u^{(0)}_2)} 
\label{schroback}
\end{equation}
where $W(f,g) = f g' - g f'$ denotes the Wronskian of $f$ and $g$. It 
follows that the second-order SUSY partner potentials $V_2(x)$ have no 
added singularities if $W(u^{(0)}_1,u^{(0)}_2)$ has no zeros.

The spectrum of $H_2$, ${\rm Sp}(H_2)$, will differ from ${\rm Sp}(H_0)$ 
depending on the normalizability of the two mathematical eigenfunctions 
$\psi^{(2)}_{\epsilon_1}$, $\psi^{(2)}_{\epsilon_2}$ of $H_2$ associated 
to $\epsilon_1$ and $\epsilon_2$ which belong to the kernel of $B$:
$$
B \psi^{(2)}_{\epsilon_j} = 0
\qquad H_2 \psi^{(2)}_{\epsilon_j} = \epsilon_j 
\psi^{(2)}_{\epsilon_j} \qquad  j=1,2
$$
For the solution associated to $\epsilon_1$ we explicitly have
\begin{eqnarray}
&& {\psi^{(2)}_{\epsilon_1}}'' + \eta {\psi^{(2)}_{\epsilon_1}}' + 
(\gamma + \eta')\psi^{(2)}_{\epsilon_1} = 0 \label{kernelf1} \\
&& {\psi^{(2)}_{\epsilon_1}}'' = 2(V_2 - \epsilon_1)\psi^{(2)}_{\epsilon_1} 
\label{hkernel1}
\end{eqnarray}
By substituting (\ref{hkernel1}) in (\ref{kernelf1}) one finds:
\begin{equation}
\eta {\psi^{(2)}_{\epsilon_1}}' + (\gamma + \eta' + 2 V_2 -2 \epsilon_1) 
\psi^{(2)}_{\epsilon_1} = 0 \label{kernelf1fo}
\end{equation}
and using the expressions for $V_2$ and $\gamma$ given in (\ref{2.23}) and 
(\ref{gamma}) with $d = \epsilon_1 + \epsilon_2$ we get:
\begin{equation}
\frac{{\psi^{(2)}_{\epsilon_1}}'}{\psi^{(2)}_{\epsilon_1}} = \frac{\eta'/2 
- \eta^2/2 + \epsilon_1 - \epsilon_2}{\eta}
\end{equation}
But from our Ans\"atz
\begin{equation}
\eta^2 = - \eta' + 2 \beta_1 \eta +2(\epsilon_1 - \epsilon_2)
\end{equation}
so that
\begin{equation}
\frac{{\psi^{(2)}_{\epsilon_1}}'}{\psi^{(2)}_{\epsilon_1}} = 
\frac{\eta'}{\eta} - \beta_1 = \frac{\eta'}{\eta} - 
\frac{{u_1^{(0)}}'}{u_1^{(0)}} \label{dlpsie12}
\end{equation}
Therefore
\begin{equation}
\psi^{(2)}_{\epsilon_1} \propto \frac{\eta}{u_1^{(0)}} \propto 
\frac{u^{(0)}_2}{W(u^{(0)}_1,u^{(0)}_2)}
\end{equation}
A similar procedure leads to
$$
\psi^{(2)}_{\epsilon_2}\propto \frac{u^{(0)}_1}{W(u^{(0)}_1,u^{(0)}_2)}
$$

Concerning the possibilities for manipulating spectra offered by the 
second-order supersymmetric quantum mechanics, we have found a heuristic 
criterion providing some interesting information. Remember that
$$
B_2 B_2^\dagger = (H_0-\epsilon_1)(H_0 - \epsilon_2)
$$
which has to be positive definite on ${\cal H}$. In particular, this has 
to be valid on the basis of energy eigenstates $\vert\psi_n^{0}\rangle$ of 
$H_0$, and thus we have
$$
\langle\psi_n^{0}\vert B_2 B_2^\dagger\vert\psi_n^{0}\rangle = 
(E_n-\epsilon_1)(E_n - \epsilon_2) \geq 0 \ \forall \ n 
$$
This opens unexpected possibilities for the positions of the new levels 
$\epsilon_1, \ \epsilon_2$. A non exhaustive list of several interesting 
situations useful for the spectral design is presented next.

\begin{itemize}

\item[$a)$] Our heuristic criterion indicates that if $\epsilon_2 
<\epsilon_1 < E_0$ it is possible to find $u^{(0)}_1$ and $u^{(0)}_2$ such 
that $W(u^{(0)}_1,u^{(0)}_2)$ is nodeless and $\psi^{(2)}_{\epsilon_1}$, 
$\psi^{(2)}_{\epsilon_2}$ are normalizable. Indeed, with this ordering of 
$\epsilon_1$ and $\epsilon_2$ the right choice is to take $u^{(0)}_1$ 
nodeless and $u^{(0)}_2$ having one zero at $x_0$, i.e., 
$u^{(0)}_2(x_0)=0$. Since $W'(u^{(0)}_1,u^{(0)}_2) = 2(\epsilon_1 - 
\epsilon_2) u^{(0)}_1u^{(0)}_2$, it turns out that 
$W(u^{(0)}_1,u^{(0)}_2)$ has one critical point at $x_0$. Moreover, we 
have that at this point $W''(u^{(0)}_1,u^{(0)}_2)/W(u^{(0)}_1,u^{(0)}_2) = 
2(\epsilon_1 - \epsilon_2) > 0$ which implies that the Wronskian acquires 
either a minimum positive or a maximum negative value at $x_0$. Hence, 
$W(u^{(0)}_1,u^{(0)}_2)$ is nodeless. The spectrum of the new Hamiltonian 
is ${\rm Sp}(H_2) = \{\epsilon_2,\epsilon_1,E_0, E_1,\dots\}$ (see a 
potential $V_2(x)$ with this kind of spectrum in figure 1). Notice that 
this case coincides with the one typically discussed when the second-order 
transformation is achieved through the iteration of two first-order 
transformations.

\begin{figure}
\resizebox{.5\textwidth}{!}
{\includegraphics{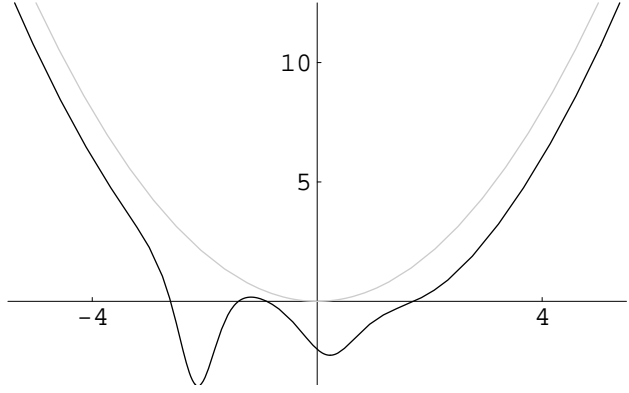}}\centering
\caption{Second-order SUSY partner potential $V_2(x)$ (black 
curve) of the harmonic oscillator (gray curve) generated by using two 
Schr\"odinger solutions $u^{(0)}_2, \ u^{(0)}_1$ of the form (\ref{gral}) 
with $\epsilon_2 = -1.2, \ \nu_2=1.1$ and $\epsilon_1 = -1, \ \nu_1=0.9$ 
respectively. Here ${\rm Sp}(H_2)=\{-1.2,-1,\frac12,\frac32,\dots\}$}
\end{figure}

\item[$b)$] If we choose $E_i<\epsilon_2<\epsilon_1<E_{i+1}$, our 
heuristic criterion suggests that perhaps we can find $u^{(0)}_1$ and 
$u^{(0)}_2$ such that $W(u^{(0)}_1,u^{(0)}_2)$ is nodeless and 
$\psi^{(2)}_{\epsilon_1}$, $\psi^{(2)}_{\epsilon_2}$ are normalizable 
\cite{fg04,sa99,fmrs02a,fmrs02b}. We would get then that ${\rm Sp}(H_2) = 
\{E_0,\dots, E_{i},\epsilon_2,\epsilon_1,E_{i+1},\dots\}$ (see figure 2). 
This possibility becomes true if we select $u^{(0)}_2$ and $u^{(0)}_1$ 
having $i+2$ and $i+1$ nodes respectively. Indeed, due to the oscillatory 
theorem, which can be translated as the fact that between two zeros of 
$u^{(0)}_2$ there is at least one zero of $u^{(0)}_1$, it turns out that 
these $2i+3$ nodes are alternating. These zeros, ordered as 
$x_0<x_1<\cdots <x_{2i+2}$, are as well the critical points of 
$W(u^{(0)}_1,u^{(0)}_2)\equiv W(x)$. Since $W(x_j)/W(x_{j+1})>0$, then 
$W(x)$ conserves its sign (i.e., it is nodeless) in the interval $(x_j, 
x_{j+1})$. Therefore, it does not have zeros in the domain 
$(x_0,x_{2i+2})$. Finally, since $x_0$ is a node of $u^{(0)}_2$, then 
$W''(x_0)/W(x_0)=2(\epsilon_1 - \epsilon_2) >0$. Hence, $W(x)$ acquires 
either a maximum negative or a minimum positive value at $x_0$. In both 
cases $W(x)$ never crosses the $x$-axis in the interval $(-\infty,x_0)$. A 
similar treatment leads us to conclude that the Wronskian does not vanish 
in $(x_{2i+2},\infty)$, and therefore it is nodeless in the full real 
line.

\begin{figure}    
\resizebox{.5\textwidth}{!}
{\includegraphics{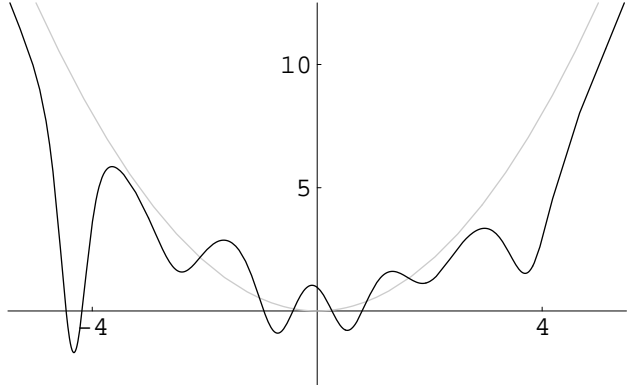}}\centering
\caption{Second-order SUSY partner potential $V_2(x)$ (black curve) of the 
oscillator (gray curve)  generated through two solutions $u^{(0)}_2, \ 
u^{(0)}_1$ of the form (\ref{gral}) with $\epsilon_2 = 3, \ \nu_2=0.9$ and 
$\epsilon_1 = 3.2,\ \nu_1=1.1$ respectively. We have that ${\rm 
Sp}(H_2)=\{\frac12,\frac32,\frac52,3,3.2,\frac72,\dots\}$.}
\end{figure}

\item[$c)$] If $\epsilon_2 = E_i$, $\epsilon_1 = E_{i+1}$, and 
$u^{(0)}_2=\psi^{(0)}_i$, $u^{(0)}_1=\psi^{(0)}_{i+1}$, then 
$W(u^{(0)}_1,u^{(0)}_2)$ is nodeless but $\psi^{(2)}_{\epsilon_1}$, 
$\psi^{(2)}_{\epsilon_2}$ are non-normalizable. In order to prove this, 
let us notice that $u^{(0)}_2$ and $u^{(0)}_1$ have now $i$ and $i+1$ 
nodes respectively. Due to the null asymptotic behavior at $x\rightarrow 
\pm\infty$ of both $u^{(0)}_1, \ u^{(0)}_2$, it turns out that these 
$2i+1$ zeros, ordered as $x_0 < x_1<\cdots < x_{2i}$, are once again 
alternating, with $x_0$ and $x_{2i}$ being now zeros of $u^{(0)}_1$. Using 
a reasoning similar to that used in the previous case, $W(x)$ becomes 
nodeless in $(x_0,x_{2i})$. On the other hand, $W(x)$ is monotonic in the 
interval $(-\infty,x_0)$ and $W''(x_0)/W(x_0)=2(\epsilon_2-\epsilon_1)<0$, 
which implies that $W(x)$ has either a maximum positive or a minimum 
negative value at $x=x_0$. Since $\lim_{x\rightarrow -\infty}W(x) = 0$, it 
turns out that the only node which $W(x)$ has in the interval 
$(-\infty,x_0)$ is an asymptotic zero as $x\rightarrow -\infty$. A similar 
procedure shows that in $(x_{2i},\infty)$ $W(x)$ has an asymptotic zero as 
$x\rightarrow \infty$. In conclusion, the Wronskian is nodeless in the 
full real line, except by asymptotic null behaviors at $x \rightarrow \pm 
\infty$. This implies that the second-order SUSY transformation is 
non-singular inside the initial domain of definition. As the intertwining 
operator respect the boundary conditions of the eigenfunctions of $H_0$ 
(except that now the eigenfunctions of $H_2$ associated to $E_i$ and 
$E_{i+1}$ are not square-integrable anymore), it turns out that ${\rm 
Sp}(H_2) = \{E_0,\dots, E_{i-1},E_{i+2},\dots\}$, i.e., somehow we have 
`deleted' the two levels $E_i$, $E_{i+1}$ in order to generate $V_2(x)$.

\end{itemize}

According to the standard SUSY treatment, for which the new levels are 
always below the ground state energy of the initial Hamiltonian, the 
previous cases (b) and (c) are somehow unexpected, supplying us with more 
freedom for manipulating spectra. In principle these non-typical cases can 
be achieved through first-order SUSY transformations, but the 
corresponding interpretation is strange: in the first step we generate a 
singular potential $V_1$, with the corresponding singularities induced by 
the zeros of the transformation function which is employed. The second 
transformation removes then the singularities introduced in the first step 
to arrive at a final non-singular potential $V_2$. Next we will explore 
another interesting cases having unexpected positions for the new levels.

\bigskip

\subsubsection{The confluent case with $c=0$} 

In this case $\xi=0$, therefore $\epsilon \equiv\epsilon_1= 
\epsilon_2\in{\mathbb R}$. Once we have found a Riccati solution 
$\beta(x)$ to (\ref{2.40}) for the given $\epsilon$, we must solve the 
Bernoulli equation resulting from (\ref{2.34}) \cite{mnr00,fs03}
$$
\eta' = - \eta^2 + 2 \beta\eta
$$
In order to solve it, let us take $\eta = 1/y$. Then
$$
y' + 2 \beta y = 1
$$
Hence
$$
y = \left[w_0 +\int e^{2\int\beta(x)dx}dx\right] e^{-2\int\beta(x)dx}
$$
$w_0$ being a real constant. Thus, the general $\eta$-solution is given by
$$
\eta(x) = \frac{e^{2\int\beta(x)dx}}{w_0 +
\int e^{2\int\beta(x)dx}dx}  
$$
In terms of the Schr\"odinger solution 
$u^{(0)}(x)\propto\exp[\int\beta(x)\,dx]$ we have
$$
\eta(x) = \frac{[u^{(0)}(x)]^2}{w_0 + \int_{x_0}^x [u^{(0)}(y)]^2\,dy} 
= \frac{w'(x)}{w(x)}
$$
where $x_0$ is a fixed point in the domain of $V_0(x)$ and, up to an 
unimportant constant factor:
\begin{equation}
w(x) = w_0 + \int_{x_0}^x [u^{(0)}(y)]^2\,dy
\label{wconfluente}
\end{equation}
In order that $V_2(x)$ will not have singularities, $w(x)$ must be 
nodeless. Since $w(x)$ is a non-decreasing monotonic function, a simple 
choice \cite{fs03} (see also \cite{am80}) is to use Schr\"odinger 
solutions such that
\begin{equation}
\lim_{x\rightarrow\infty} u^{(0)}(x) = 0 \ \ {\rm and} \ \ 
I_+ = \int_{x_0}^\infty [u^{(0)}(y)]^2\,dy <\infty  \label{inftyplus}
\end{equation}
or 
\begin{equation}
\hskip-0.15cm
\lim_{x\rightarrow-\infty} u^{(0)}(x) = 0 \ \ {\rm and} \ \
I_- = \int_{-\infty}^{x_0} [u^{(0)}(y)]^2\,dy <\infty
\label{inftyminus}
\end{equation}
In both cases it is possible to find a $w_0$-domain where $w(x)$ is 
nodeless \cite{fs03}. For instance, if (\ref{inftyplus}) is valid and 
$u^{(0)}(x)$ is a non-physical eigenfunction of $H_0$ associated to 
$\epsilon$, it turns out that $\lim_{x\rightarrow-\infty} w(x) = -\infty$ 
and $\lim_{x\rightarrow\infty} w(x) = w_0 + I_+$. Thus, the domain for 
which $w(x)$ is nodeless is $w_0\leq -I_+$. A similar procedure implies 
that for non-physical transformation functions satisfying 
(\ref{inftyminus}) the nodeless $w_0$-domain is $w_0 \geq I_-$.

Similarly as in the case with $c>0$, it can be found now a function 
$\psi^{(2)}_{\epsilon}$ in the kernel of $B$ which is simultaneously an 
eigenfunction of $H_2$ with eigenvalue $\epsilon$. Indeed, let us notice 
that equations (\ref{kernelf1}-\ref{dlpsie12}) remain valid in this case, 
we just have to substitute $\epsilon_1$ and $\epsilon_2$ by $\epsilon$, 
$u_1^{(0)}$ by $u^{(0)}$ and $\beta_1$ by $\beta$. Thus:
$$
\psi^{(2)}_{\epsilon}(x) \propto \frac{\eta(x)}{u^{(0)}(x)} =  
\frac{u^{(0)}(x)}{w(x)} \label{missing}
$$

\begin{figure}    
\resizebox{.5\textwidth}{!}
{\includegraphics{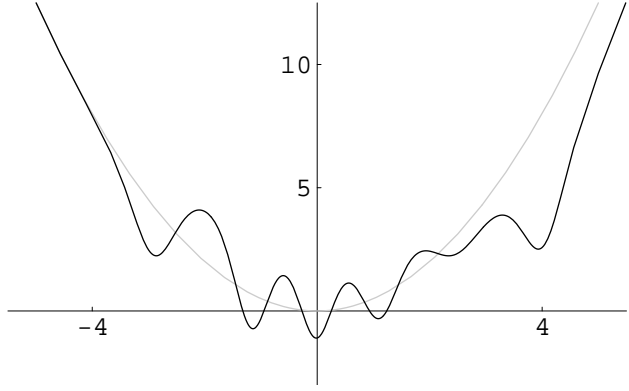}}\centering 
\caption{Confluent second-order SUSY partner potential $V_2(x)$ (black 
curve) of the oscillator (gray curve) generated by using a solution 
$u^{(0)}$ of the form (\ref{gral}) with $\epsilon=4, \ \nu=1$ and taking 
$w_0=5, \ x_0=0$ in (\ref{wconfluente}).  Now ${\rm Sp}( 
H_2)=\{\frac12,\frac32,\frac52,\frac72,4,\frac92,\dots\}$.}
\end{figure}

The spectrum of $H_2$ depends on the normalizability of 
$\psi^{(2)}_{\epsilon}$. In particular, for $\epsilon\geq E_0$ it is 
possible to find solutions $u^{(0)}$ satisfying (\ref{inftyplus}) or 
(\ref{inftyminus}) and such that $\psi^{(2)}_{\epsilon}$ is normalizable. 
This means that the confluent second-order SUSY transformations allow the 
embedding of {\it single} energy levels above the ground state energy of 
$H_0$ (see figure 3). In addition, the physical solutions associated to 
the excited state levels of $H_0$ can be used as well as transformation 
functions, no matter the number of zeros they have in the domain of 
$V_0(x)$ (see figure 4). These two features cannot be achieved by 
iterating the first-order SUSY transformations without paying the price of 
introducing singularities at the intermediate potentials. As expected, 
these atypical spectral possibilities are consistent with our heuristic 
criterion previously formulated.

\begin{figure}    
\resizebox{.5\textwidth}{!}
{\includegraphics{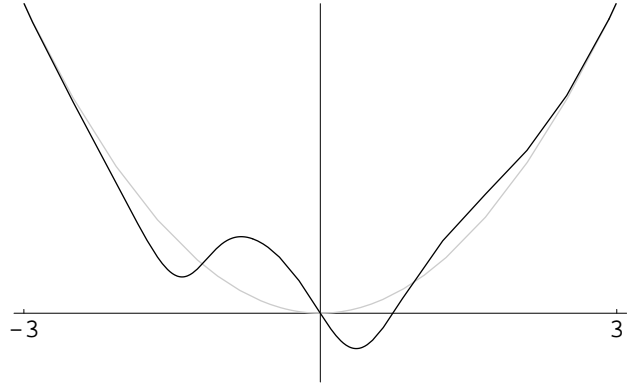}}\centering
\caption{Confluent isospectral second-order SUSY partner potential 
$V_2(x)$ (black curve) of the oscillator (gray curve) generated by using 
the first excited state $u^{(0)} = \psi^{(0)}_1$ of the oscillator, with 
$\epsilon=E_1=\frac32$, $w_0=1$ and $x_0=0$.}
\end{figure}

\subsubsection{The complex case with $c<0$} 

If $c<0$ we have that $\epsilon_1 \in {\mathbb C}$ and 
$\epsilon_2=\bar\epsilon_1$. Notice that our heuristic criterion allows 
this possibility without violating the positive nature of the operator 
$B_2B_2^\dagger$. We are going to analize just the case for which $V_2(x)$ 
is real valued, implying that $\beta_2(x) = \bar\beta_1(x)$. Following 
analogous steps as for the real case, one arrives to the solution 
$\eta(x)$ of (\ref{2.33}) in terms of the (complex) solution $\beta_1(x)$ 
of the Riccati equation (\ref{2.40}) associated to $\epsilon_1$ 
\cite{fmr03} (for the case when $V_2(x)$ is allowed to be complex see 
\cite{rm03,mu04}):
\begin{equation}
\eta(x)= -\frac{2{\rm Im}(\epsilon_1)}{{\rm Im}[\beta_1(x)]}
\end{equation}
Using the corresponding complex Schr\"o\-din\-ger solution $u^{(0)}_1(x)$ 
we can write
$$
\eta(x) = \frac{w'(x)}{w(x)} \qquad\quad
w(x) = \frac{W(u^{(0)}_1,\bar u^{(0)}_1)}{2(\epsilon_1 - \bar \epsilon_1)}
$$

\begin{figure}    
\resizebox{.5\textwidth}{!}
{\includegraphics{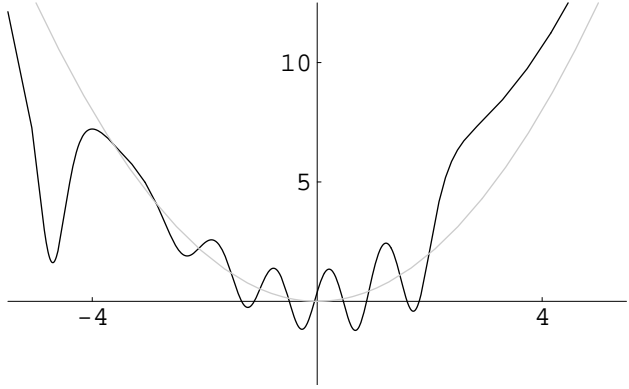}}\centering
\caption{Real isospectral second-order SUSY partner potential $V_2(x)$ 
(black curve) of the oscillator (gray curve) ge\-ne\-ra\-ted by using a 
solution $u^{(0)}_1$ of the form (\ref{gral}) with $\nu_1=-1$ and 
$\epsilon_1 = 5.4 +i/20\in{\mathbb C}$.}
\end{figure}

In order to avoid the creation of singularities in $ V_2(x)$, $w(x)$ must 
be nodeless. Since $w'(x) = \vert u^{(0)}_1(x)\vert^2$, it turns out that 
$w(x)$ is a non-decreasing monotonic function. Thus, to ensure that 
$w(x)\neq 0 \ \forall \ x\in {\mathbb R}$ it is sufficient that 
\cite{fmr03}
\begin{equation}
\lim_{x\rightarrow\infty} u^{(0)}_1(x) = 0 \ \ {\rm or} \ \
\lim_{x\rightarrow-\infty} u^{(0)}_1(x) = 0 \label{complexlimit}
\end{equation}
For transformation functions obeying the condition (\ref{complexlimit}) it 
turns out that $V_2(x)$ is a real potential isospectral to $V_0(x)$ (see 
figure 5). Let us notice that, as in the previous cases, this 
transformation could be achieved by iterating the first-order SUSY, but 
the intermediate potentials would be complex. This is, perhaps, the reason 
why this case was almost unexplored in the past. From our viewpoint, 
however, it represents a very promising line of research for the next 
years (see e.g. \cite{rami03}).

\newpage

\section{An example: the harmonic oscillator}

We are going to apply the previous techniques to the harmonic oscillator. 
Before doing that, however, it is convenient to discuss some generalities 
of a subset of nonlinear deformations of the Heisenberg algebra. We will 
realize later on the importance of such structures for the SUSY partners 
of the oscillator.

\subsection{Polynomial deformations of the Heisenberg algebra}

First of all let us remember the standard oscillator algebra
\begin{eqnarray*} 
& [H,a^\dagger] = a^\dagger \qquad [H,a] =  - a \\
& [a,a^\dagger] = 1
\end{eqnarray*}
for which the number operator is a linear function of $H$:
\begin{eqnarray*}
& N=a^\dagger a = H - \frac12
\end{eqnarray*}

On the other hand, the polynomial Heisenberg algebras of $m$-th order are 
deformations of the previous structure, where there are two standard 
commutation relationships
\begin{equation}
[H,L^\pm] = \pm L^\pm \label{commlie}
\end{equation}
and an atypical one characterizing the deformation:
\begin{equation}
[L^-,L^+] \equiv N(H+1) - N(H)=P_m(H)
\label{commdef}
\end{equation}
where a generalization of the standard number operator is given by 
$N(H)\equiv L^+ L^-$. The corresponding systems are described by 
Schr\"odinger Hamiltonians
\begin{equation}
H = -\frac12 \frac{d^2}{dx^2} + V(x) \label{hamiltonian}
\end{equation}
being $L^\pm$ differential ladder operators of order $(m+1)$-th, $N(H)$ a 
polynomial of order $(m+1)$-th in $H$ factorized as
\begin{equation}
N(H) = \prod_{i=1}^{m+1} \left(H - {\cal E}_i\right)
\label{nfactorized}
\end{equation}
and $P_m(H)$ a $m$-th order polynomial in $H$. The algebra 
(\ref{commlie}-\ref{nfactorized}) generated by $\{H,L^-,L^+\}$ provides 
information on the spectrum ${\rm Sp}(H)$ of $H$ 
\cite{fh99,dek92,qv99,acin00}. Indeed, let us consider the solution space 
of the $(m+1)$-th order differential equation (the kernel $K_{L^-}$ of 
$L^-$):
\begin{equation}
L^-\,\psi = 0 \quad \Rightarrow \quad
L^+ L^-\,\psi = \prod_{i=1}^{m+1}\left(H - {\cal E}_i\right)\psi = 0
\label{kernell-}
\end{equation}
Notice that $K_{L^-}$ is invariant under $H$:
$$
L^- H \psi = (H+1)L^-\psi = 0 \ \ \Rightarrow \ \ H\psi \in K_{L^-} \ 
\forall \  \psi \in K_{L^-}
$$
Thus, it is natural to select as the basis of $K_{L^-}$ the functions 
which are simultaneously eigenstates of $H$ with eigenvalues ${\cal E}_i$
\begin{equation}
H\psi_{{\cal E}_i} = {\cal E}_i \psi_{{\cal E}_i}
\end{equation}
These represent the {\it extremal\/} states for the $m+1$ {\it 
mathematical ladders\/} of spacing $\Delta E = 1$ starting from ${\cal 
E}_i$. If $s$ of these states are physically meaningful, $\{\psi_{{\cal 
E}_i}, i = 1,\dots,s\}$, then by acting iteratively with $L^+$ onto them 
$s$ physical energy ladders can be constructed (see figure 6-a).

\begin{figure}    
\resizebox{0.9\textwidth}{!}
{\includegraphics{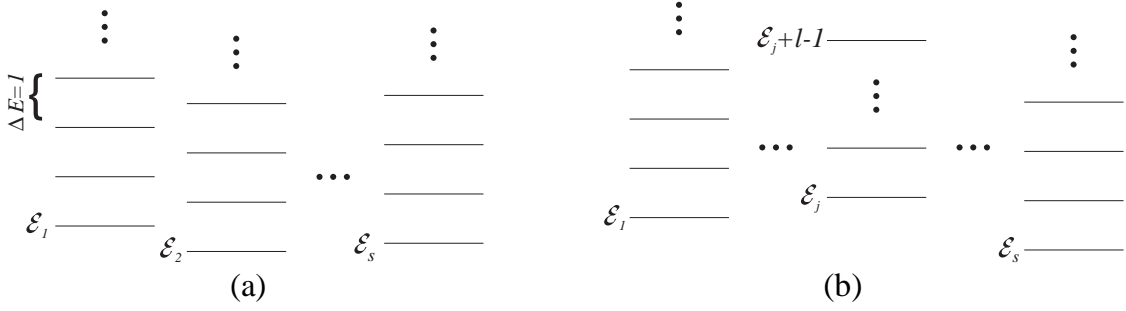}}\centering
\caption{Possible spectra for the Hamiltonians satisfying 
(\ref{commlie})--(\ref{nfactorized}) with $s$ physical extremal states. In 
case (a) $s$ infinite ladders are obtained by acting with $L^+$. In case 
(b) there are $s-1$ infinite ladders, and a finite one (the $j$-th one) is 
built out of $\psi_{{\cal E}_j}$ taking into account 
(\ref{finiteladderfig2}).}
\end{figure}

It could happen \cite{fh99} that for the ladder starting from ${\cal E}_j$ 
there is a $l\in {\mathbb N}$ such that:
\begin{equation}
\label{finiteladderfig2}
\left( L^+\right)^{l-1}\psi_{{\cal E}_j} \neq 0 \qquad \left(
L^+\right)^{l}\psi_{{\cal E}_j} = 0
\end{equation}
Then
\begin{eqnarray*}
&& L^- (L^+)^{l}\psi_{{\cal E}_j} =  L^-L^+ (L^+)^{l-1}\psi_{{\cal 
E}_j} \\ && = \left(\prod_{i=1}^{m+1} \left(H + 1 - {\cal 
E}_i\right)\right)
(L^+)^{l-1}\psi_{{\cal E}_j} \\ && =  \left(\prod_{i=1}^{m+1} \left({\cal 
E}_j + l - {\cal E}_i\right) \right)(L^+)^{l-1}\psi_{{\cal E}_j} = 0
\end{eqnarray*}
This implies that another root of (\ref{nfactorized}) must have the form 
${\cal E}_k = {\cal E}_j + l$, $k\in\{s+1,\dots,m+1\}, j\in 
\{1,\dots,s\}$.  Hence, ${\rm Sp}(H)$ will contain $s-1$ infinite ladders 
and a finite one of length $l$, starting from ${\cal E}_j$ and ending at 
${\cal E}_j+l-1$ (see figure 6-b).

We conclude that the spectrum of systems described by polynomial 
Heisenberg algebras of order $m$ can have at most $m+1$ infinite ladders. 
Notice that pairs of ladder operators for the harmonic oscillator 
satisfying (\ref{commlie}-\ref{nfactorized}) with $m>0$ can be constructed 
simply by taking $L^-=a P(H)$, $L^+=P(H)a^\dagger$, where $a,a^{\dagger}$ 
are the annihilation and creation operators of the oscillator and $P(H)$ 
is a real polynomial in $H$ \cite{dgrs99}. These deformations are 
reducible since for the same system we already have ladder operators $a, 
a^{\dagger}$ obeying a much simpler algebra. Here we will be mainly 
interested in the search of irreducible deformed algebras.

\subsection{SUSY partners of the oscillator}

Let us consider the harmonic oscillator potential
$$
V_0(x) = \frac{x^2}2
$$
The corresponding Hamiltonian has an equidistant spectrum with 
eigenfunctions and eigenvalues given by:
\begin{equation}
\psi_n^{(0)}(x) = \sqrt{\frac{1}{2^nn!\sqrt{\pi}}} \, H_n(x) 
e^{-\frac{x^2}2}
\qquad E_n = n + \frac12, \ n=0,1,\dots
\end{equation}
where $H_n(x)$ are the Hermite polynomials. In order to implement the 
several SUSY transformations discussed previously, we will solve the 
Schr\"odinger equation (\ref{schrodinger}) for an arbitrary factorization 
energy $\epsilon$, namely:
\begin{eqnarray}
&& -\frac12 {u^{(0)}}'' + \frac{x^2}2 u^{(0)} = \epsilon u^{(0)} \nonumber
\end{eqnarray}
Suppose that $u^{(0)} = e^{-\frac{x^2}2} h(x)$. Therefore:
$$
h'' -2 x h' + (2\epsilon-1)h = 0
$$
By changing variables $y=x^2$ we arrive at:
\begin{equation}
y \frac{d^2h}{dy^2} + (b-y) \frac{dh}{dy}-a h=0 \label{che}
\end{equation}
which is the confluent hypergeometric equation with $a=(1-2\epsilon)/4$ 
and $b=1/2$. The general solution of (\ref{che}) leads to the general 
solution $u^{(0)}$ we were looking for:
\begin{eqnarray}
&& u^{(0)}(x) = e^{-\frac{x^2}2}
\bigg[{}_1F_1\left(a,\frac12;x^2\right) + 2\nu x\frac{\Gamma(a + 
\frac12)}{\Gamma(a)} 
\, {}_1 F_1\left(a + \frac12,\frac32;x^2\right)\bigg]
\label{gral}
\end{eqnarray}
where $\epsilon\in{\mathbb C}$ and ${}_1F_1\left(a,b;y\right)$ is the 
confluent hypergeometric (Kummer) function:
\begin{equation}
{}_1F_1\left(a,b;y\right) = \frac{\Gamma(b)}{\Gamma(a)} 
\sum_{n=0}^{\infty}\frac{\Gamma(a+n)}{\Gamma(b+n)}\frac{y^n}{n!}
\end{equation}
Notice that, for $\epsilon<1/2$ and $\vert\nu\vert\leq1$ the solution 
$u^{(0)}$ given in (\ref{gral}) is nodeless.

The explicit expressions for the SUSY partners of the oscillator can be 
calculated using (\ref{gral}) and, in general, they are too involved to be 
shown here. However, we will present the simplest derivation of a family 
of potentials isospectral to the oscillator \cite{mi84}, which originally 
was derived by Abraham and Moses through the Gelfand-Levitan formalism 
\cite{am80}.

\subsubsection{The Abraham-Moses potentials}

Let us make a first-order SUSY transformation employing the general 
Schr\"odinger solution (\ref{gral}) associated to $\epsilon = -1/2$, 
namely:
\begin{eqnarray*}
& u^{(0)}(x) = e^{\frac{x^2}2}[1 + \nu {\rm Erf}(x)]
\end{eqnarray*}
where
\begin{eqnarray*}
{\rm Erf}(x) = \frac{2}{\sqrt{\pi}}\int_0^x e^{-t^2}dt
\end{eqnarray*}
is the well known error function. The corresponding superpotential is 
given by
$$
\alpha_1(x,\epsilon) = x + \frac{2\nu e^{-\frac{x^2}2}}{\sqrt{\pi}[1 +
\nu {\rm Erf}(x)]}
$$
which is precisely the one obtained by Mielnik through his generalized 
factorization \cite{mi84}. The first-order SUSY partner potentials are 
given by:
\begin{equation}
V_1(x) = \frac{x^2}2 - \left(\frac{2\nu e^{-\frac{x^2}2}}{\sqrt{\pi}[1 +
\nu {\rm Erf}(x)]}\right)' - 1 \label{amp}
\end{equation}
Up to a displacement in the energy origin, this is the Abraham-Moses 
family of potentials isospectral to the oscillator \cite{am80}. A plot of 
these potentials in terms of the variable $x$ for continuous values of the 
parameter $\nu$ in $[-0.95,0]$ is given in figure 7. Notice that for 
$\nu=0$ we recover the original oscillator potential.

\begin{figure}    
\resizebox{.5\textwidth}{!}
{\includegraphics{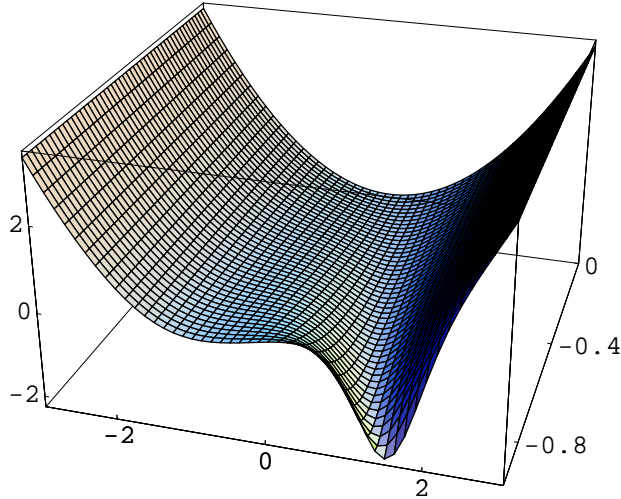}}\centering
\caption{The Abraham-Moses family of potentials (\ref{amp}) as a function 
of $x$ and continuous values of the $\nu$-parameter in the domain 
$[-0.95,0]$.}
\end{figure}

Let us perform now a confluent second-order SUSY transformation employing 
the ground state eigenfunction, namely:
\begin{equation}
u^{(0)} =  \frac{e^{-\frac{x^2}2}}{\pi^{1/4}} \qquad \epsilon = \frac12 
\label{gso}
\end{equation}
For the confluent algorithm the key $w$-function becomes (see equation 
(\ref{wconfluente})):
\begin{eqnarray*}
& w(x) = w_0 + \frac12 {\rm Erf}(x)
\end{eqnarray*}
where we have made $x_0=0$. Taking now $1/w_0 =2\nu$, the new potentials 
read:
\begin{eqnarray*}
V_2(x) & = & \frac{x^2}2 - \left(\frac{w'}{w}\right)' = 
\frac{x^2}2 - \left(\frac{2\nu e^{-\frac{x^2}2}}{\sqrt{\pi}[1 +
\nu {\rm Erf}(x)]}\right)'
\end{eqnarray*}
These are precisely the already mentioned Abraham-Moses potentials 
(compare with (\ref{amp})). Indeed, the procedure of derivation based in 
the confluent algorithm employing (\ref{gso}) essentially coincides with 
the Abraham-Moses treatment \cite{am80}. Although in this case the 
first-order SUSY and the confluent second-order SUSY algorithms led to the 
same family of potentials, this was a fortunate coincidence due to the 
huge symmetry of the oscillator. In the general case, the potentials 
derived through the first-order SUSY will not coincide with those 
generated through second-order SUSY transformations.

\subsection{Non-linear algebra of $H_k$}

Suppose now that we have applied a $k$-th order SUSY transformation to 
$V_0 =x^2/2$, producing then the potential $V_k$ by creating $k$ new 
energy levels. The spectrum of the end Hamiltonian $H_k$, intertwined with 
the harmonic oscillator through $B_k^\dagger$, will be $\{\epsilon_i,E_n = 
n+1/2, \ i=k,\dots,1,\ n=0,1, \dots\}$, i.e., it contains a part 
isospectral to the oscillator plus $k$ additional levels $\epsilon_i, \ 
i=k,\dots,1$ placed by simplicity below $E_0=1/2$. Hence, it is possible 
that polynomial Heisenberg algebras rule the H-SUSY partners of the 
oscillator.

To analyze the algebraic structure characteristic of the Hamiltonians 
$H_k$, let us look for ladder operators which connect the eigenstates 
associated to the levels $E_n$. There is a natural construction for a pair 
of these operators \cite{mi84,fh99,fhn94,fnr95}, which is guessed from 
equation (\ref{intertwiningk}), its adjoint and the standard intertwining 
relationship involving the oscillator Hamiltonian $H_0$ and its creation 
and annihilation operators $a^\dagger$, $a$:
\begin{equation}
(H_0-1)a^\dagger = a^\dagger H_0 \qquad (H_0+1) a = a H_0 
\end{equation}
The construction is composed of three stages (see figure 8): i) first we 
`move' the eigenvectors $\vert\psi_n^k\rangle$ of $H_k$, represented 
previously by the wave functions $\psi_n^{(k)}$, to the eigenvectors 
$\vert\psi_n^0\rangle$ of the oscillator Hamiltonian $H_0$ through the 
intertwining operator $B_k$. ii) Then, we move up 
($\vert\psi_{n+1}^0\rangle$) or down ($\vert\psi_{n-1}^0\rangle$) on the 
ladder of $H_0$ by using $a^\dagger$ or $a$ respectively, which will cause 
the effective `motion' up or down on the ladder of $H_k$. iii)  Finally, 
we get back to the ladder of $H_k$ by acting $B_k^\dagger$ on 
$\vert\psi_{n+1}^0\rangle$ or $\vert\psi_{n-1}^0\rangle$. Thus, the 
`natural' ladder operators for $H_k$ are:
\begin{equation}
L_k = B_k^\dagger a B_k \qquad L_k^\dagger = B_k^\dagger a^\dagger B_k
\qquad k=0,1,\dots \label{naturallo}
\end{equation}
For completeness, we have extended the intertwining relationship 
(\ref{intertwiningk}) to the case $k=0$ by assuming that $B_0^\dagger = 
B_0 = I$, $I$ is the identity operator. The action of $L_k$ and 
$L_k^\dagger$ is drawn just onto the spectral points $E_n=n+1/2, \
n=0,1,\dots$ because the eigenstates $\{ \vert\psi_{\epsilon_i}^k\rangle, 
\ i=1,\dots,k\}$ are annihilated by both $L_k$ and $L_k^\dagger$ since 
they are annihilated by $B_k$.

\begin{figure}    
\resizebox{.6\textwidth}{!}
{\includegraphics{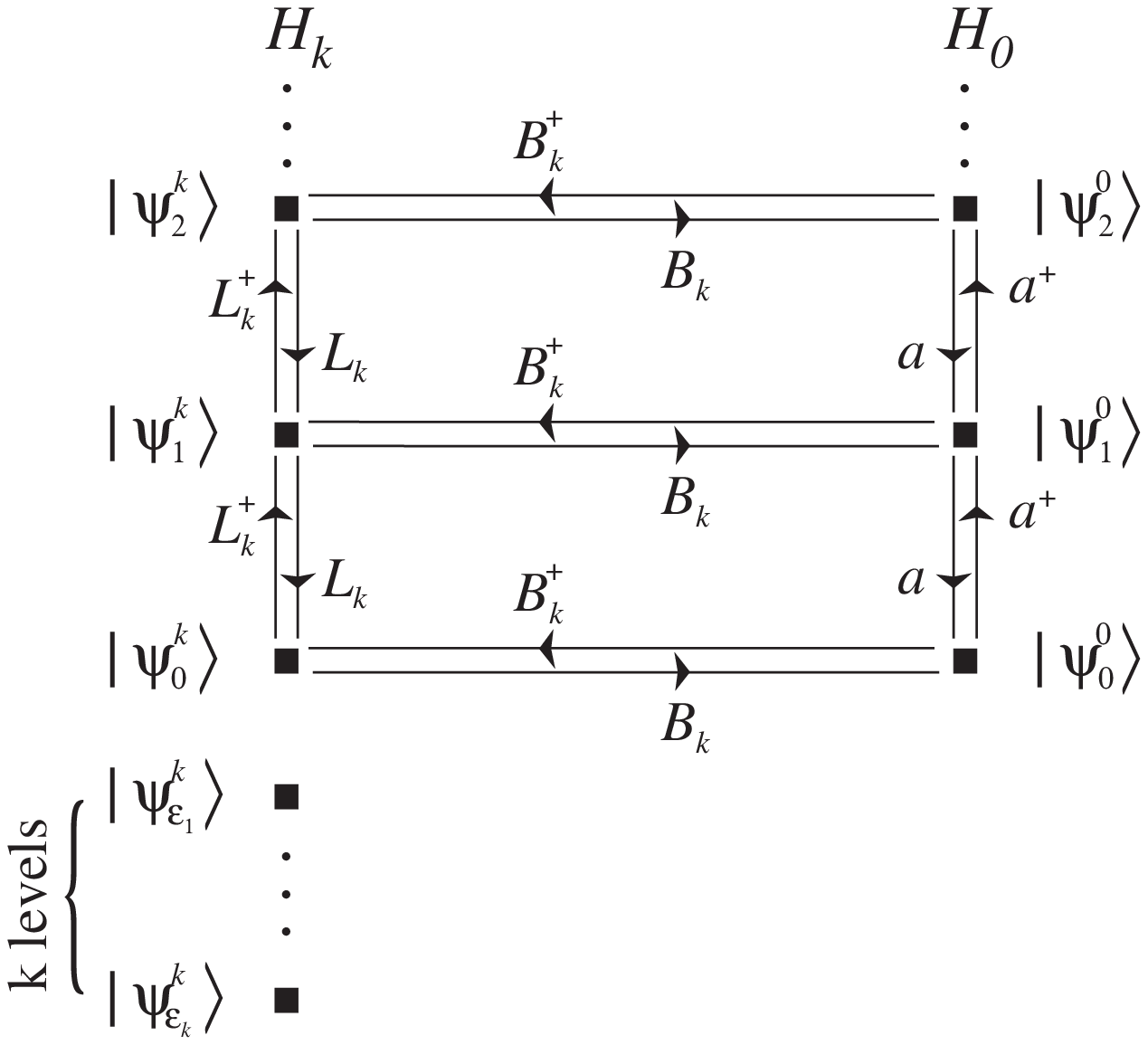}}\centering
\caption{Diagram representing the action of the $k$-th order intertwining 
operators $B_k, \ B_k^\dagger$ and of the ladder operators $a, \ 
a^\dagger, \ L_k, \ L_k^\dagger$ for the Hamiltonians $H_0$ and $H_k$}
\end{figure}

Notice that
\begin{eqnarray*}
H_k L_k & = & H_k B_k^\dagger a B_k = B_k^\dagger  H_0 a B_k =B_k^\dagger 
a(H_0-1) B_k \\
& = & B_k^\dagger a B_k (H_k-1) = L_k (H_k-1)
\end{eqnarray*}
i.e., $L_k$ and $L_k^\dagger$ are differential ladder operators of order 
$(2k+1)$-th satisfying:
\begin{equation}
[H_k,L_k]=-L_k \qquad [H_k,L_k^\dagger] = L_k^\dagger 
\end{equation}
The operator $N(H_k) \equiv L_k^\dagger L_k$, which generalizes the 
standard number operator for the harmonic oscillator, is a polynomial in 
$H_k$ of $(2k+1)$-th order \cite{dek92,ek95,vs93}):
\begin{eqnarray}
N(H_k) & \equiv & L_k^\dagger L_k = B_k^\dagger a^\dagger B_k
B_k^\dagger a B_k = B_k^\dagger a^\dagger \prod_{i=1}^k \left( H_0 - 
\epsilon_i\right) a B_k \nonumber \\ 
&  = &  B_k^\dagger \left(H_0-\frac12\right) 
\prod_{i=1}^k
\left( H_0 - \epsilon_i - 1\right)B_k \nonumber \\
& = & \left( H_k - \frac12\right) 
\prod_{i=1}^k \left( H_k - \epsilon_i - 1\right) \left( H_k - \epsilon_i 
\right)
\label{ngen}
\end{eqnarray}
Thus, the operators $L_k, \ L_k^\dagger$ and $H_k$ generate a polynomial 
Heisenberg algebra of order $2k$:
\begin{equation}
[L_k,L_k^\dagger] =N(H_k+1) -  N(H_k) \label{conmeles}
\end{equation}
For consistency, when $k=0$ we should get the standard Heisenberg algebra 
because $L_0=a$ and $L_0^\dagger = a^\dagger$. This linear case is indeed 
recovered from our formulae:
\begin{equation} 
[H_0,L_0] = -L_0 \qquad [H_0,L_0^\dagger] = L_0^\dagger \qquad [L_0, 
L_0^\dagger] = I  
\end{equation} 
The corresponding number operator becomes the standard linear expression 
in terms of the oscillator Hamiltonian:
\begin{eqnarray} 
& N(H_0) = H_0 - \frac12 = N
\end{eqnarray} 
On the other hand, when $k=1$ and $\epsilon_1$ is arbitrary we recover the 
quadratic Heisenberg algebra \cite{as97} (see also \cite{dek92,ek95}):
\begin{equation} 
[L_1,L_1^\dagger] = (H_1 -\epsilon_1)(3 H_1 - \epsilon_1)
\end{equation} 
The number operator becomes now cubic in $H_1$: 
\begin{eqnarray} 
& N(H_1) = \left(H_1 - \frac12\right)\left(H_1
-\epsilon_1\right) \left(H_1 -\epsilon_1 - 1\right)
\end{eqnarray} 
If $k=2$ we will get a polynomial Heisenberg algebra of fourth order:
\begin{equation}
[L_2,L_2^\dagger] = (H_2 - \epsilon_1)(H_2 - \epsilon_2)\left[5 H_2^2 -
3(\epsilon_1 + \epsilon_2)  H_2 + \epsilon_1\epsilon_2 + 1 \right]
\end{equation} 
where $N(H_2)$ is a $5$-th order polynomial of $H_2$:
\begin{eqnarray} 
& N(H_2) =
\left(H_2 - \frac12\right)\left(H_2 -\epsilon_1\right) \left(H_2
-\epsilon_2\right)  \left(H_2 -\epsilon_1 - 1\right) \left(H_2 
-\epsilon_2 - 1\right)
\end{eqnarray} 
For general $k$, it arises a polynomial Heisenberg algebra of order $2k$
whose properties are characterized by the $(2k+1)$-th order polynomial
$N(H_k)$ of (\ref{ngen}).

It should be clear now why the roots of (\ref{ngen}) are 
$\{1/2,\epsilon_i, \epsilon_i+1, i=1,\dots,k\}$: our H-SUSY partner 
Hamiltonians $H_k$ have precisely $k+1$ physical extremal states 
associated to the $k+1$ roots (eigenvalues) $\{1/2,\epsilon_i, 
i=1,\dots,k\}$. As the ladder starting from $1/2$ is infinite, it does not 
imply any restriction on the remaining roots of the $(2k+1)$-th order 
polynomial $N(H_k)$. However, as the ladders starting from $\epsilon_i$ 
are of length $1$ (they end again at the initial energy $\epsilon_i$), the 
other $k$ roots become $\epsilon_i + 1, i=1,\dots,k$, as those appearing 
in (\ref{ngen}).

An interesting point to be addressed next, concerning the non-linear 
nature of the polynomial algebras (\ref{naturallo}-\ref{conmeles}), is 
that they can be partially linearized \cite{fh99,fnr95,ro96}.

\subsection{Linearization of the non-linear algebra of $H_k$}

As pointed out previously, the $k$ eigenstates of $H_k$, 
$\vert\psi_{\epsilon_i}^k\rangle, i=1,\dots,k$, are isolated between 
themselves and from the $\vert\psi_n^k\rangle$ because 
$L_k\vert\psi_{\epsilon_i}^k\rangle = 0 = \ 
L_k^\dagger\vert\psi_{\epsilon_i}^k\rangle$. Hence, it seems natural to 
look for a linearization on the subspace spanned by 
$\{\vert\psi_n^{k}\rangle, n=0,1,\dots\}$. The method consists in 
modifying the ladder operators $L_k$ and $L_k^\dagger$ of 
(\ref{naturallo}) to get an action similar to the resulting one when the 
Heisenberg algebra generators are applied to the appropriate energy 
eigenstates \cite{fh99,fnr95,ro96}. As for the sub-basis 
$\{\vert\psi_n^k\rangle, n=0,1,\dots\}$ the commutator $[L_k,L_k^\dagger]$ 
is already diagonal (see equation (\ref{conmeles})), we propose a 
modification which will convert most of the diagonal elements of 
$[L_k,L_k^\dagger]$ to $1$, namely:
\begin{equation} 
L_L = B_k^\dagger f(N) a B_k \qquad L_L^\dagger = B_k^\dagger a^\dagger
f(N) B_k 
\end{equation} 
$N = a^\dagger a$ being the standard number operator for the harmonic 
oscillator, $f(x)$ a real function to be determined and the subscript $L$ 
denoting linearization. We ask that $[L_L,L_L^\dagger] = I$ on the 
subspace ${\cal H}_{\geq 1}$ spanned by $\{\vert \psi_n^k\rangle, 
n=1,2,\dots\}$. Notice that we leave open the possibility that 
$[L_L,L_L^\dagger]\vert \psi_0^k\rangle = {\rm c} \vert \psi_0^k\rangle, \ 
{\rm c}\neq 1 \in {\mathbb R}$. It is straightforward to show that
\begin{eqnarray*}
& L_L \vert \psi_n^k\rangle = \sqrt{g(n)} \vert \psi_{n-1}^k\rangle 
\qquad
L_L^\dagger \vert \psi_n^k\rangle = \sqrt{g(n+1)} \vert
\psi_{n+1}^k\rangle \\
& [L_L,L_L^\dagger] \vert\psi_n^k\rangle = [g(n+1) - g(n)]\vert\psi_n^k
\rangle
\end{eqnarray*}
where
\begin{eqnarray}
& g(n) = n\left[f(n-1)\right]^2 \prod_{i=1}^k 
\left(n-\epsilon_i-\frac12\right)
\left(n-\epsilon_i+\frac12\right)  \\
& g(n+1)-g(n) = 1  \qquad n = 1,2,\dots \label{fde}
\end{eqnarray}
The general solution of the finite difference equation (\ref{fde}) 
becomes:
\begin{equation}
g(n) = n+{\rm w}(n)
\end{equation}
${\rm w}(n)$ being periodic with period $1$, ${\rm w}(n+1) = {\rm w}(n), 
\ n=1,2,\dots$ Hence:
\begin{equation}
f(n-1) = \sqrt{\frac{n+{\rm w}(n)}{n
\prod_{i=1}^k (n-\epsilon_i-\frac12)(n-\epsilon_i+\frac12)}}
\end{equation}
Since ${\rm w}(n)$ takes the same value for all $n=1,2,\dots$, it is 
important just ${\rm w}\equiv {\rm w}(1)$. Moreover, $f(n-1)$ should be 
real which implies that ${\rm w}\geq -1$. Finally, the ladder operators we 
were looking for read \cite{fh99}:
\begin{eqnarray}
&& L_L = B_k^\dagger \sqrt{\frac{N+1+{\rm w}}{(N+1)
\prod_{i=1}^k (N-\epsilon_i+\frac12)(N-\epsilon_i+\frac32)}}\, a B_k
\label{anopl} \\
&& L_L^\dagger = B_k^\dagger a^\dagger \sqrt{\frac{N+1+{\rm w}}{(N+1)
\prod_{i=1}^k (N-\epsilon_i+\frac12)(N-\epsilon_i+\frac32)}}\, B_k 
\label{cropl}
\end{eqnarray}

Although their explicit forms are more involved than the ones for $L_k$ 
and $L_k^\dagger$ (compare (\ref{naturallo}) with 
(\ref{anopl},\ref{cropl})), however $L_L$ and $L_L^\dagger$ act simpler on 
the energy eigenstates $\vert\psi_n^k\rangle, n=0,1,\dots$ (except by the 
case with $k=0$ which is separately discussed):
\begin{eqnarray}
& L_L \vert\psi_n^k\rangle = (1 - \delta_{n0})\sqrt{n+{\rm w}} \,\vert
\psi_{n-1}^k\rangle \qquad L_L^\dagger \vert\psi_n^k\rangle = 
\sqrt{n+{\rm w}+1} 
\, \vert\psi_{n+1}^k \rangle \\
& [L_L,L_L^\dagger] \vert\psi_n^k\rangle = (1 +
{\rm w}\delta_{n0})\vert\psi_{n}^k\rangle 
\end{eqnarray}
This representation is independent of $k$, i.e., of the order of the 
interwining operator used to go from $H_0$ to $H_k$. The modified algebra 
here derived coincides with the `distorted' Heisenberg algebra originally 
introduced to linearize partially the second-order Heisenberg algebra 
characterizing the Abraham-Moses potentials, where ${\rm w}\geq -1$ is the 
distortion parameter \cite{fnr95} (see also \cite{fh99}). We have shown 
that this distorted algebra is common to all the H-SUSY partners of the 
oscillator.  Moreover, a `complete' linearization on ${\cal H}_{\geq 0}$ 
(the subspace spanned by $\{\vert\psi_n^k\rangle, n = 0,1,\dots\}$) can be 
achieved by taking ${\rm w}=0$ to obtain precisely the Heisenberg algebra 
representation, namely:
\begin{equation}
L_L\vert\psi_n^k\rangle = \sqrt{n} \vert\psi_{n-1}^k\rangle \qquad
L_L^\dagger\vert\psi_n^k\rangle = \sqrt{n+1} \vert\psi_{n+1}^k\rangle
\qquad
[L_L,L_L^\dagger]\vert\psi_n^k\rangle = \vert\psi_n^k\rangle
\end{equation}
If ${\rm w}=-1$ we get once again the standard Heisenberg algebra on 
${\cal H}_{\geq 1}$ but the state $\vert\psi_0^k\rangle$ is annihilated by 
both $L_L$ and $L_L^\dagger$, i.e., it has been isolated {\it by hand} of 
the rest of eigenstates of $H_k$. This isolation property is {\it natural} 
for the other $k$ eigenstates $\vert \psi_{\epsilon_i}^k\rangle, i = 1, 
\dots, k$.

Notice that for $k=0$ the explicit expressions for $L_L$, $L_L^\dagger$ do 
not coincide with $a$, $a^\dagger$:
\begin{equation}
L_L= \sqrt{\frac{N+1+{\rm w}}{N+1}}\,a \qquad L_L^\dagger = a^\dagger
\sqrt{\frac{N+1+{\rm w}}{N+1}}
\end{equation}
Now we get a distortion of the Heisenberg algebra representation which 
maps the operators $a, \ a^\dagger$ into $L_L, \ L_L^\dagger$, by changing 
the matrix elements of $a$ and $a^\dagger$ without affecting the diagonal 
elements of $[a,a^\dagger]$ except the one associated to 
$\vert\psi_0^0\rangle$, which becomes ${\rm w} + 1$. For ${\rm w} = 0$ we 
recover the original Heisenberg algebra because now $L_L=a, \ 
L_L^\dagger=a^\dagger$. Moreover, when ${\rm w} = -1$ we get a reducible 
representation composed of the Heisenberg algebra on ${\cal H}_{\geq 1}$ 
and the null representation on the subspace generated by 
$\vert\psi_0^0\rangle$ due to 
$L_L\vert\psi_0^0\rangle=L_L^\dagger\vert\psi_0^0\rangle = 0$.

\newpage

\section{Coherent states for the SUSY partners of the oscillator}

The beautiful properties of the harmonic oscillator coherent states (CS) 
motivated the interest in looking for them in other physical situations 
\cite{fhn94,fnr95,ro96,fa93,sp95,kk96,fs96,bs96,ek97,sbl98,cjt98, 
bg71,zfg80,ks85,pe86,bd89,chks95,gn96,mwv99,sps99,spk01,do02}. There are 
various definitions, each one of them leading in general to different sets 
of CS. Concerning the intertwining technique, CS which are eigenstates of 
a certain annihilation operator for the shape invariant potentials were 
constructed by Fukui and Aizawa \cite{fa93}. The CS as eigenstates of the 
annihilation operator $L_k$ of (\ref{naturallo}) with $k=1$ and $\epsilon 
= -1/2$, i.e. for the Abraham-Moses family of isospectral oscillator 
potentials were derived in 1994 \cite{fhn94}, and the linearization 
process in the same case as well as the corresponding CS analysis was 
elaborated in \cite{fnr95,ro96}. Since then, a lot of works have arisen 
looking for interrelations between CS and quantum groups, 
pseudodifferential operators, non-linear algebras, etcetera 
\cite{sp95,kk96,fs96,bs96,ek97,sbl98,cjt98,sps99,spk01}. In particular, 
the CS construction for the H-SUSY partners of the oscillator with 
arbitrary $k$ has been successfully addressed \cite{fh99}. For didactic 
purposes, let us discuss first the main properties of the standard CS 
\cite{bg71,zfg80,ks85,pe86,bd89}.

\subsection{Standard coherent states}

There are three equivalent definitions of the harmonic oscillator coherent 
states which can be used to define them in other physical situations.

\begin{enumerate}

\item The coherent states $\vert z\rangle$ are eigenstates of the 
annihilation operator
\begin{equation}
a \vert z\rangle = z\vert z\rangle \qquad z \in {\mathbb C} \label{eigena}
\end{equation}

\item The coherent states $\vert z\rangle$ are obtained by applying the 
displacement operator $D(z)$ onto the oscillator ground state $\vert 
0\rangle$ (in this subsection we use the standard notation (Fock) for the 
eigenstates of the number operator, i.e., $\vert n\rangle \equiv 
\vert\psi_n^0\rangle$):
\begin{equation}
\vert z\rangle = D(z)\vert 0\rangle \qquad D(z) = e^{za^\dagger - 
\bar z a}
\end{equation}

\item The coherent states $\vert z\rangle$ are quantum states with a 
minimum-uncertainty relationship
\begin{eqnarray}
& (\Delta x)(\Delta p)=\frac12
\end{eqnarray}
where for a system in the state $\vert z\rangle$ and an arbitrary operator 
${\cal O}$ the corresponding uncertainty expresses:
\begin{equation}
\Delta{\cal O} = \sqrt{\langle z \vert ({\cal O} - \langle z \vert {\cal 
O} \vert z \rangle)^2 \vert z \rangle} = \sqrt{\langle z \vert {\cal 
O}^2\vert z \rangle - \langle z \vert {\cal O}\vert z \rangle^2}
\end{equation}
\end{enumerate}

Since for general systems we are specially interested in the first 
definition, we derive here the standard CS through (\ref{eigena}). First 
we expand $\vert z\rangle$ in the Fock basis
\begin{equation}
\vert z\rangle = \sum_{n=0}^\infty c_n \vert n\rangle
\end{equation}
Then we use (\ref{eigena}) to obtain:
\begin{eqnarray*}
a\vert z\rangle & = & \sum_{n=1}^\infty c_n \sqrt{n}\vert n-1\rangle = 
\sum_{n=0}^\infty z c_n \vert n\rangle =
\sum_{n=1}^\infty z c_{n-1} \vert n-1\rangle
\end{eqnarray*}
Hence, we get a recursion formula for $c_n$:
\begin{equation}
c_n = \frac{zc_{n-1}}{\sqrt{n}}
\end{equation}
By iterating this equation we arrive at:
\begin{equation}
c_n = \frac{z^nc_0}{\sqrt{n!}}
\end{equation}
Finally, by asking that $c_0$ is a positive constant such that $\langle 
z\vert z\rangle =1$ we get:
\begin{equation}
\vert z\rangle = e^{-\frac{r^2}2}\sum_{n=0}^\infty
\frac{z^n}{\sqrt{n!}} 
\vert n\rangle
\end{equation}
where $r=\vert z\vert$. Let us remark that the standard coherent states 
can be used as an alternative basis (non-orthogonal) in the Hilbert space 
of states because they form a complete set:
\begin{eqnarray}
& \frac{1}{\pi}\int \vert z\rangle \langle z \vert d^2z = I
\end{eqnarray}
This can be straightforwardly verified since in the polar representation 
$z = re^{i\theta}$ we have:
\begin{eqnarray}
\frac{1}{\pi}\int \vert z\rangle \langle z \vert d^2z & = & 
\frac{1}{\pi} 
\sum_{n,m=0}^{\infty}\frac{\vert n\rangle\langle m\vert}{\sqrt{n!m!}} 
\int_0^\infty e^{-r^2} r^{n+m+1}dr \int_0^{2\pi} e^{i\theta(n-m)}d\theta 
\nonumber \\
& = & \sum_{n=0}^{\infty}\frac{\vert n\rangle\langle n\vert}{n!} 
\ 2\int_0^\infty e^{-r^2}r^{2n+1}dr = \sum_{n=0}^{\infty}\vert 
n\rangle\langle n\vert = I
\end{eqnarray}
Thus, any state can be expanded in the basis of coherent states. In 
particular, any CS $\vert z'\rangle$ admits a non-trivial decomposition:
\begin{eqnarray}
& \vert z'\rangle  = \frac{1}{\pi}\int \vert z\rangle \langle z \vert 
z'\rangle  d^2z
\end{eqnarray}
where the reproducing Kernel $\langle z \vert z'\rangle$ is given by
\begin{equation}
\langle z \vert z'\rangle = e^{-\frac{r^2}{2} + \bar z z' - 
\frac{r'^2}{2}}
\end{equation}
Let us notice that coherent states evolve into coherent states, namely
\begin{eqnarray*}
e^{-itH} \vert z\rangle & = & e^{-\frac{r^2}2}\sum_{n=0}^\infty
\frac{z^n}{\sqrt{n!}} e^{-i(n+\frac12)t}\vert n\rangle = 
e^{-i\frac{t}2}\vert z(t)\rangle \qquad z(t) = ze^{-it}
\end{eqnarray*}

\subsection{Coherent states for $H_k$}

Let us construct the CS for the SUSY partner Hamiltonians $H_k$ of the 
oscillator as eigenstates of the annihilation operators $L_k$ and $L_L$. 
First, let us find them as eigenstates of $L_k$ (the non-linear case):
\begin{equation}
L_k\vert z\rangle = z\vert z \rangle \qquad z\in{\mathbb C} 
\label{eigenelek}
\end{equation}
We express $\vert z\rangle$ as a linear combination of the subset of 
eigenstates $\vert\psi_n^k\rangle$ of $H_k$ associated to the part of the 
spectrum isospectral to the oscillator:
\begin{equation}
\vert z\rangle = \sum_{n=0}^\infty c_n \vert\psi_n^k\rangle
\label{expansionelek}
\end{equation}
After inserting (\ref{expansionelek}) in (\ref{eigenelek}) and using the 
fact that
\begin{equation}
L_k \vert\psi_n^k\rangle = \sqrt{n
\prod_{i=1}^k \left(n-\epsilon_i - \frac12\right)\left(n - \epsilon_i + 
\frac12\right)} 
\ \vert\psi_{n-1}^k\rangle
\end{equation}
we get a recurrence relationship for the coefficients $c_n$
\begin{eqnarray}
c_{n}= \frac{z c_{n-1}}{\sqrt{n
\prod_{i=1}^k (n-\epsilon_i - \frac12)(n-\epsilon_i + \frac12)}}
= \frac{\sqrt{\prod_{i=1}^k \Gamma(-\epsilon_i + 
\frac12)\Gamma(-\epsilon_i +
\frac32)}z^n c_{0}}{\sqrt{n!
\prod_{i=1}^k \Gamma(n\!-\!\epsilon_i\!+\!\frac12)\Gamma(n\!-\!\epsilon_i
\!+\! \frac32)}}
\end{eqnarray}
We fix $c_0$ by the condition $\langle z\vert z\rangle = 1$ and asking 
that $c_0>0$. Hence, our non-linear CS become:
\begin{eqnarray}
& \vert z\rangle = 
\sum_{n=0}^\infty \frac{\sqrt{
\prod_{i=1}^k \Gamma(-\epsilon_i + \frac12)\Gamma(-\epsilon_i + \frac32)
}\ z^n \vert\psi_n^k\rangle}{\sqrt{
n! {}_0\!F_{2k}(   
-\epsilon_1 + \frac12,\dots,-\epsilon_k + \frac12,-\epsilon_1 +
\frac32,\dots
,-\epsilon_k + \frac32;r^2)
\prod_{i=1}^k \Gamma(n - \epsilon_i + \frac12)\Gamma(n - \epsilon_i
+ \frac32)}}\label{nlcs}
\end{eqnarray}
being $\Gamma(x)$ the gamma function, $r = \vert z\vert$, and ${}_pF_q$ a 
generalized hypergeometric function:
\begin{equation}
{}_pF_q(a_1,\dots,a_p,b_1,\dots,b_q;x) = \frac{\Gamma(b_1) \dots
\Gamma(b_q)}{\Gamma(a_1)\dots\Gamma(a_p)} \sum_{n=0}^\infty 
\frac{\Gamma(a_1+n)\dots\Gamma(a_p+n)}{\Gamma(b_1+n) \dots \Gamma(b_q +n)}
\frac{x^n}{n!}
\end{equation}
Notice that $z=0$ is a $(k+1)$-th degenerate eigenvalue of $L_k$ because 
we get of (\ref{nlcs}) that $\vert z=0\rangle = \vert\psi_0^k\rangle$ and 
$L_k\vert\psi_{\epsilon_i}^k\rangle = 0, \ i=1,\dots,k$. Thus, the 
resolution of the identity should be looked for as:
\begin{equation}
I = \sum_{i=1}^k \vert\psi_{\epsilon_i}^k\rangle\langle
\psi_{\epsilon_i}^k\vert + \int \vert z\rangle\langle z\vert d\mu(z)
\label{idres}
\end{equation}
where the measure $d\mu(z)$ is to be determined. Suppose now that
\begin{eqnarray}
& d\mu(z) = {}_0F_{2k}\left(-\epsilon_1+\frac12, \dots, 
-\epsilon_k+\frac12,
-\epsilon_1 + \frac32, \dots, -\epsilon_k + \frac32;r^2\right) h(r^2) r dr
d\theta
\end{eqnarray}
Inserting this equation in (\ref{idres}) and using the fact that $\{\vert
\psi_{\epsilon_i}^k\rangle, \vert\psi_n^k\rangle, \ i=1,\dots,k, \
n=0,1,\dots\}$ is complete, we arrive at the following requirement for
$h(x)$:
\begin{equation}
\int_0^\infty x^n h(x)dx = \frac{\Gamma(n+1)
\prod_{i=1}^k\Gamma(n-\epsilon_i+\frac12)\Gamma(n-\epsilon_i +
\frac32)}{\pi\prod_{i=1}^k\Gamma(-\epsilon_i+\frac12)\Gamma(-\epsilon_i
+\frac32)} \label{mellin}
\end{equation}
Hence, $h(x)$ is the inverse Mellin transform of the right hand side of 
(\ref{mellin}) \cite{ma83}. It turns out that $h(x)$ is proportional to a 
Meijer $G$-function \cite{fh99,ma83,ba54}:
\begin{equation}
h(x) = \frac{ G^{2k+1 \ \ 0}_{\ \ 0 \ \ 2k+1}(x\vert
0,-\epsilon_1-\frac12,\dots
,-\epsilon_k-\frac12,-\epsilon_1+\frac12,\dots,-\epsilon_k+\frac12)} 
{\pi \prod_{i=1}^k\Gamma(-\epsilon_i+\frac12)\Gamma(-\epsilon_i   
+\frac32)} \label{meijer}
\end{equation}
Let us notice that for $k=1$ and $\epsilon_1=-1/2$ an explicit expression 
for (\ref{meijer}) has been derived in \cite{fhn94}.

The other properties of the standard coherent states have their 
corresponding analogue here. For instance, any CS of the form (\ref{nlcs}) 
can be expressed in terms of the others:
\begin{equation}
\vert z'\rangle = \int \vert z\rangle\langle z\vert z'\rangle d\mu(z)
\end{equation}
where the reproducing Kernel $\langle z\vert z'\rangle$ can be 
straightforwardly evaluated:
\begin{eqnarray}
& \hskip-1cm \langle z\vert z'\rangle  \! = \!  
\frac{{}_0F_{2k}(-\epsilon_1 + \frac12, \dots,
-\epsilon_k + \frac12, -\epsilon_1 + \frac32, \dots, -\epsilon_k +
\frac32;{\bar z}z')}
{\sqrt{{}_0F_{2k}(-\epsilon_1 + \frac12,\dots,
-\epsilon_k  + \frac12, -\epsilon_1 + \frac32,\dots,
-\epsilon_k + \frac32;r^2) 
{}_0F_{2k}(-\epsilon_1 + \frac12,\dots,
-\epsilon_k  + \frac12, -\epsilon_1 +
\frac32,\dots,-\epsilon_k + \frac32 ; r'^2)}}  
\end{eqnarray}
This means that any two CS $\vert z\rangle$ and $\vert z'\rangle$ of form 
(\ref{nlcs}) are non-orthogonal. From the resolution of the identity 
(\ref{idres}) it is clear that any state vector can be expanded in terms 
of our CS if we include the atypical orthogonal CS 
$\vert\psi_{\epsilon_i}^k\rangle, \ i=1,\dots, k$ naturally inherent to 
this treatment. Notice also that our CS evolve in time as coherent states:
\begin{eqnarray*}
U(t) \vert z\rangle = e^{-\frac{it}2} \vert z(t)\rangle \qquad z(t) = 
ze^{-it}
\end{eqnarray*}

Let us derive now the coherent states for the linearized annihilation 
operator $L_L$ of (\ref{anopl}) as:
\begin{equation}
L_L \vert z,{\rm w}\rangle = z \vert z,{\rm w}\rangle \label{eigenelel}
\end{equation}
where we have made explicit the CS dependence on the distortion parameter 
${\rm w}$. The expansion of $\vert z,{\rm w}\rangle$ in the sub-basis 
$\vert\psi_n^k\rangle$ is equal to the one arising in the right hand side 
of (\ref{expansionelek}), and the use of (\ref{eigenelel}) leads to the 
following recurrence relationship for the $c_n$'s:
\begin{eqnarray*}
c_n = \frac{z c_{n-1}}{\sqrt{n+{\rm w}}} = 
\frac{\sqrt{\Gamma({\rm w}+1)}
z^n}{\Gamma(n+{\rm w} + 1)} c_0
\end{eqnarray*}
The coefficient $c_0$ is determined by asking that $\langle z, {\rm 
w}\vert z,{\rm w}\rangle = 1$ and $c_0>0$. Finally
\begin{equation}
\vert z,{\rm w}\rangle = \sqrt{\frac{\Gamma({\rm 
w}+1)}{{}_1F_1(1,{\rm w} + 1;r^2)}}
\sum_{n=0}^\infty \frac{z^n}{\sqrt{\Gamma(n+{\rm 
w} + 1)}}\vert\psi_n^k\rangle
\label{csl}
\end{equation}
The resolution of the identity becomes similar to (\ref{idres}):
\begin{equation}
I= \sum_{i=1}^k \vert\psi_{\epsilon_i}^k\rangle\langle
\psi_{\epsilon_i}^k\vert + \int \vert z, {\rm w}\rangle\langle z, 
{\rm w}\vert
d\mu_L(z)
\end{equation}
where the measure $d\mu_L(z)$ is given by:
\begin{equation}
d\mu_L(z) =\sigma(r,{\rm w}) \,r\,dr\,d\!\theta \qquad \sigma(r,{\rm w}) =
\frac{{}_1F_1(1,{\rm w} + 1;r^2)}{\pi\Gamma({\rm w} + 1)} e^{-r^2} 
r^{2{\rm w}}
\end{equation}
The reproducing Kernel is now
\begin{equation}
\langle z,{\rm w}\vert z',{\rm w}\rangle = \frac{{}_1F_1(1, {\rm 
w} + 1;{\bar z}z')}{\sqrt
{{}_1F_1(1,{\rm w}+1;r^2){}_1F_1(1,{\rm w}+1;r'^2)}}
\end{equation}
For ${\rm w} = 0$ (the linearized case in ${\cal H}_{\geq 0}$) the same 
formulae as for the standard coherent states are recovered by noticing 
that ${}_1F_1(1,1;r^2) = e^{r^2}$. Moreover, by taking carefully the limit 
${\rm w}\rightarrow -1$ (the linearized case now in ${\cal H}_{\geq 1}$) 
it can be shown that the standard expression for the CS is also recovered, 
but the eigenstate $\vert\psi_0^k\rangle$ associated to the eigenvalue 
$E_0=1/2$ will be isolated of the other ones, i.e., the series (\ref{csl}) 
will start from $\vert\psi_1^k\rangle$ \cite{fh99,fnr95,ro96}.

A comparison of the annihilation operators $L_k$ and $L_L$ and of both 
sets of coherent states derived in this section shows the following: the 
explicit expression for the non-linear operator $L_k$ is simpler than the 
one for $L_L$.  As can be seen from equations (\ref{nlcs}) and 
(\ref{csl}), however, the CS associated to $L_L$ are less involved than 
the ones associated to $L_k$, which is due to the simplest algebra 
representation generated by $L_L$ and $L_L^\dagger$.

Let us analyze, for both kinds of CS, the uncertainty products $(\Delta 
x)(\Delta p)$ in cases for which we get closed analytic expressions. This 
can be directly achieved if we restrict ourselves to H-SUSY partner 
potentials in the harmonic oscillator limit. In particular, for $k=1$ and 
$\epsilon_1 = -1/2$ the uncertainty product for the non-linear CS 
(\ref{nlcs}) was calculated for the Abraham-Moses potentials in the 
oscillator limit \cite{fhn94}, i.e., by taking $\nu=0$ in (\ref{amp}). We 
shall present this case as well as the non-linear one with $k=2$, 
$\epsilon_1 = -1/2, \ \epsilon_2 = - 3/2$, and taking $\nu_1 = 0,$ 
$\nu_2\rightarrow \infty$ in order to recover the oscillator limit. For 
the linearized coherent states we shall study the case with $k=1$, 
$\epsilon_1 = -1/2$, $\nu_1=0$, ${\rm w} = 1$ (which was analyzed in 
\cite{fnr95,ro96}) as well as the one with $k=2, \ \epsilon_1 = -1/2, \ 
\epsilon_2 = - 3/2$, $\nu_1 = 0,$ $\nu_2\rightarrow \infty$, ${\rm w} = 
2$. Notice that the corresponding expressions for the CS can be obtained 
from equations (\ref{nlcs}) and (\ref{csl}) by realizing that in the 
oscillator limit $\vert\psi_n^2\rangle \rightarrow 
\vert\psi_{n+2}^0\rangle$ and $\vert\psi_n^1\rangle \rightarrow 
\vert\psi_{n+1}^0\rangle$. Henceforth, the sets of coherent states we are 
dealing with read:
\begin{eqnarray}
& \vert z\rangle  = \frac{1}{\sqrt{{}_0F_2(1,2;r^2)}}
\sum_{n=0}^\infty \frac{z^n}{n!\sqrt{(n+1)!}}\vert\psi_{n+1}^0
\rangle & \qquad k=1 \label{csnlk1} \\
& \vert z\rangle  = \sqrt{\frac{2}{{}_0F_4(1,2,2,3;r^2)}}
\sum_{n=0}^\infty \frac{z^n}{n!(n+1)!\sqrt{(n+2)!}}\vert\psi_{n+2}^0
\rangle & \qquad k=2 \label{csnlk2} \\
& \vert z,{\rm w}=1 \rangle  = 
\frac{1}{\sqrt{{}_1F_1(1,2;r^2)}}
\sum_{n=0}^\infty \frac{z^n}{\sqrt{(n+1)!}}\vert\psi_{n+1}^0
\rangle & \qquad k=1 \label{cslk1} \\
& \vert z,{\rm w} = 2 \rangle  = 
\sqrt{\frac{2}{{}_1F_1(1,3;r^2)}}
\sum_{n=0}^\infty \frac{z^n}{\sqrt{(n+2)!}}\vert\psi_{n+2}^0
\rangle & \qquad k=2 \label{cslk2}
\end{eqnarray}
A direct calculation using (\ref{csnlk1}) and $x=(a+a^\dagger)/\sqrt2$, $p 
= i(a^\dagger - a)/\sqrt2$ leads to the uncertainties $\Delta x$ and 
$\Delta p$ in the non-linear case for $k=1$:
\begin{eqnarray}
&&\Delta x = \sqrt{\frac32 - [{\rm Re}(z)]^2\rho_1(r)}  \\
&& \Delta p = \sqrt{\frac32 - [{\rm Im}(z)]^2\rho_1(r)} \\
&& \rho_1(r) = 2 \left[
\frac{{}_0F_2(2,2;r^2)}{{}_0F_2(1,2;r^2)}\right]^2 -
\frac{{}_0F_2(2,3;r^2)}{{}_0F_2(1,2;r^2)}
\end{eqnarray}
where ${\rm Re}(z)$ and ${\rm Im}(z)$ represent the real and imaginary 
parts of $z$ respectively. A plot of the corresponding uncertainty product 
is given in figure 9.

\begin{figure}    
\resizebox{.5\textwidth}{!}
{\includegraphics{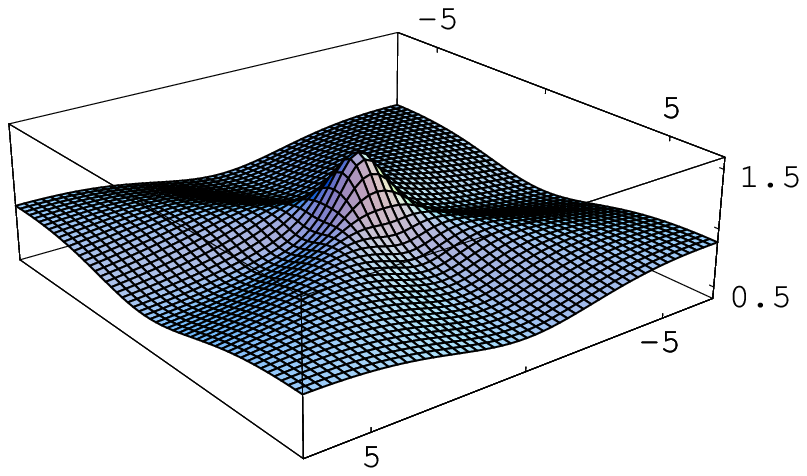}}\centering
\caption{The uncertainty product $(\Delta x)(\Delta p)$ (vertical axis) as 
function of $z$ for the non-linear coherent states (\ref{csnlk1}) 
associated to the oscillator}
\end{figure}

On the other hand, for $k=2$ and employing the non-linear CS in 
(\ref{csnlk2}) it is found that:
\begin{eqnarray}
&& \Delta x = \sqrt{\frac52 - [{\rm Re}(z)]^2\rho_2(r)} \\
&& \Delta p = \sqrt{\frac52 - [{\rm Im}(z)]^2\rho_2(r)} \\
&& \rho_2(r) = \frac12 \left[
\frac{{}_0F_4(2,2,3,3;r^2)}{{}_0F_4(1,2,2,3;r^2)}\right]^2 - \frac16 
\left[\frac{{}_0F_4(2,3,3,4;r^2)}{{}_0F_4(1,2,2,3;r^2)}\right]
\end{eqnarray}
A plot of $(\Delta x)(\Delta p)$ is given in figure 10.

\begin{figure}    
\resizebox{.5\textwidth}{!}
{\includegraphics{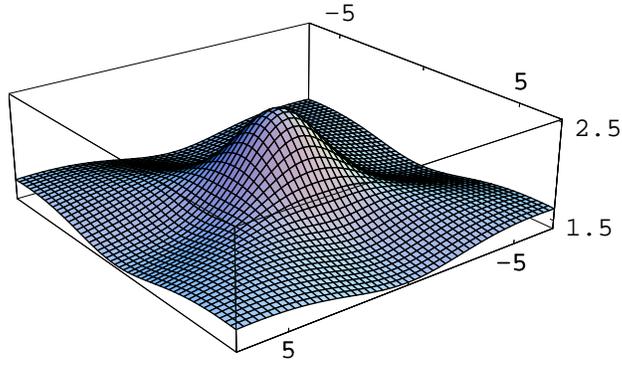}}\centering
\caption{The uncertainty product $(\Delta x)(\Delta p)$ (vertical axis) as 
function of $z$ for the non-linear coherent states (\ref{csnlk2}) 
associated to the oscillator}
\end{figure}

On the other hand, in the linearized case with $k=1$ and making use of the 
CS in (\ref{cslk1}) we get:
\begin{equation}
(\Delta x)^2 = (\Delta p)^2 = (\Delta x)(\Delta p) = \frac12 +
\frac{1}{{}_1F_1(1,2;r^2)}
\end{equation}
A plot of the uncertainty product is given in figure 11.

\begin{figure}    
\resizebox{.5\textwidth}{!}
{\includegraphics{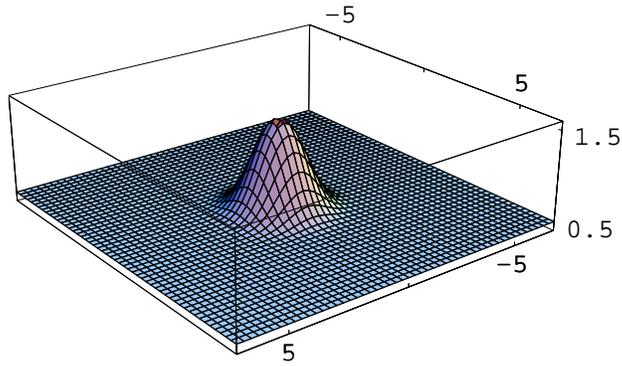}}\centering
\caption{The uncertainty product $(\Delta x)(\Delta p)$ (vertical axis) as 
function of $z$ for the linearized coherent states (\ref{cslk1}) of the 
oscillator}
\end{figure}

Finally, in the linearized case with $k=2$ and employing (\ref{cslk2}) we 
find:
\begin{equation}
(\Delta x)^2 = (\Delta p)^2 = (\Delta x)(\Delta p) = \frac12 +
\frac{2}{{}_1F_1(1,3;r^2)}
\end{equation}
A plot of $(\Delta x)(\Delta p)$ is given in figure 12. 

\begin{figure}    
\resizebox{.5\textwidth}{!}
{\includegraphics{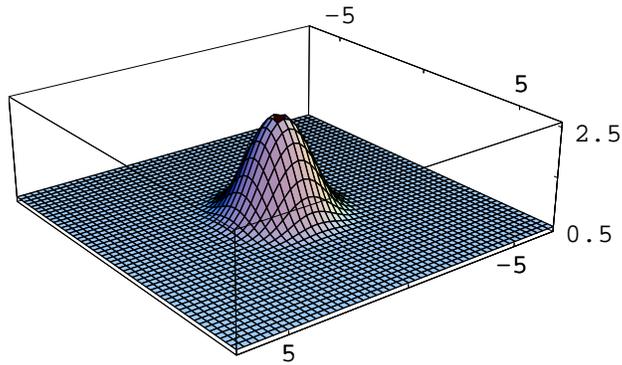}}\centering
\caption{The uncertainty product $(\Delta x)(\Delta p)$ (vertical axis) as 
function of $z$ for the linearized coherent states (\ref{cslk2}) of the 
oscillator}
\end{figure}

As we can see, $(\Delta x)(\Delta p)$ have more structure in the 
non-linear than in the linearized cases. Indeed, in the linear cases 
$(\Delta x)(\Delta p)$ have radial symmetry, they differ of the standard 
result $(\Delta x)(\Delta p)=1/2$ just in a neighborhood about the 
origin, and they quickly approach the standard value when $\vert z\vert 
\rightarrow \infty$ (see figure 11 and 12). This does not happens for the 
non-linear CS for which the asymptotic values of $(\Delta x)(\Delta p)$ 
are in general different from $1/2$ and depend on the direction in which 
we move out of $z=0$ (see figures 9 and 10).

We conclude that the annihilation and creation operators for the 
oscillator H-SUSY partner Hamiltonians which are more similar to the 
standard oscillator ones become $L_L$ and $L_L^\dagger$. They mimic quite 
well the annihilation and creation operators of the oscillator and lead to 
the standard CS expression in the cases when the distortion parameter 
takes the two values ${\rm w} = 0$ and ${\rm w} = -1$. Moreover, $L_L$ and 
$L_L^\dagger$ become exactly equal to $a$ and $a^\dagger$ when $k= {\rm w} 
= 0$, and the corresponding CS are precisely the standard ones for the 
oscillator.

\newpage

\section{Conclusions and outlook}

We have discussed the possibilities for designing quantum spectra offered 
by the supersymmetric quantum mechanics. We have seen that the standard 
iterative method, in which the first-order SUSY transformations are used 
to construct higher-order ones, could induce the wrong conclusion that the 
new Hamiltonians will have always the new levels below the ground state 
energy of the initial one. Here, it has been shown that the direct 
procedure frees us of that belief, allowing to create under certain 
restrictions some new levels above the initial ground state energy. It is 
important to underline also the complex SUSY transformations which can be 
employed to generate non-hermitian Hamiltonians with either purely real 
spectra or with some finite number of complex levels 
\cite{fmr03,rm03,mu04}. We think that this interesting line of research is 
worth to be continued.

When the SUSY techniques are applied to the harmonic oscillator, some 
interrelations immediately arise with several subjects of mathematical 
physics as non-linear deformations of Lie algebras and coherent states 
\cite{fh99}. It can be established also connections between SUSY techniques 
and non-linear ordinary differential equations as Riccati \cite{crf01}, 
Painlev\'e \cite{cfnn04} and KdV \cite{mr04}. In particular, when looking 
for the general systems ruled by second-order polynomial Heisenberg 
algebras, it turns out that the corresponding potentials become determined 
by a certain function which obeys the Painlev\'e IV equation 
\cite{acin00,vs93}. It has been recently shown that subsets of H-SUSY 
partner Hamiltonians of the oscillator for which the $k$ levels 
$\epsilon_i, \ i=k,\dots,1$ are forced to be connected by appropriate 
annihilation and creation operators to form a finite ladder of equally 
spaced levels, supply us with explicit solutions of the Painlev\'e IV 
equations \cite{fnn04}. Notice that a similar treatment relates subsets of 
SUSY partner potentials of the radial oscillator and solutions of the 
Painlev\'e V equation \cite{cfnn04}.

We conclude by mentioning that links of this kind between SUSY techniques 
and non-linear aspects of mathematical-physics, as the ones previously 
pointed out, represent the future of a field which has proved very 
fruitful along the years (see e.g. \cite{afhnns04}).

\bigskip

\noindent{\bf Acknowledgements.} The authors acknowledge the support of 
CONACYT (M\'exico), project No. 40888-F.

\end{document}